\documentclass[letterpaper,twocolumn,10pt]{article}
\usepackage{usenix-2020-09}

\usepackage{tikz}
\usepackage{amsmath}

\usepackage{xcolor,color,xspace,enumerate,centernot,multirow,float,graphicx,
xcolor,caption,subcaption,textcomp,pgfplots,pgf-pie,tikz,listings,enumitem,
comment,adjustbox,mdframed,changepage,algorithm,algorithmic}
\usepackage{hyperref}
\usepackage{filecontents}
\usepackage{xurl}

\usepackage[utf8]{inputenc}
\usetikzlibrary{shapes}
\usetikzlibrary{arrows.meta}
\usetikzlibrary{arrows}
\usetikzlibrary{positioning}
\usetikzlibrary{shadows}
\usetikzlibrary{backgrounds}
\usepackage[most]{tcolorbox}
\definecolor{lightblue}{rgb}{0.68, 0.85, 0.9}
\definecolor{lightpink}{rgb}{1.0, 0.71, 0.76}
\newenvironment{myquote}
  {\begin{adjustwidth}{0.2cm}{0.1cm}}
  {\end{adjustwidth}}

\definecolor{darkgray}{gray}{0.6}

\newmdenv[
  leftline=true,
  rightline=false,
  topline=false,
  bottomline=false,
  linecolor=darkgray,
  linewidth=2pt,
  innerleftmargin=5pt,
  innerrightmargin=0pt,
  innertopmargin=0pt,
  innerbottommargin=0pt
]{leftbar}

\newtcbox{\bluebox}[1][blue]{
  on line,
  colback=blue!5,
  colframe=blue!75!black,
  boxrule=0.1pt,
  arc=1pt,
  boxsep=0pt,left=0.5pt,right=0.5pt,top=0.5pt,bottom=0.5pt
}

\newtcbox{\lightbluebox}[1][blue]{
  on line,
  colback=lightblue,
  colframe=lightblue,
  boxrule=0.1pt,
  arc=1pt,
  boxsep=0pt,left=0.5pt,right=0.5pt,top=0.5pt,bottom=0.5pt
}

\newtcbox{\pinkbox}[1][blue]{
  on line,
  colback=pink!5,
  colframe=pink!75!black,
  boxrule=0.1pt,
  arc=1pt,
  boxsep=0pt,left=0.5pt,right=0.5pt,top=0.5pt,bottom=0.5pt
}

\newtcbox{\lightpinkbox}[1][blue]{
  on line,
  colback=lightpink,
  colframe=lightpink,
  boxrule=0.1pt,
  arc=1pt,
  boxsep=0pt,left=0.5pt,right=0.5pt,top=0.5pt,bottom=0.5pt
}

\newcommand{\Closed}[0]{Closed\xspace}
\newcommand{\CookieWait}[0]{Cookie\_Wait\xspace}
\newcommand{\CookieEchoed}[0]{Cookie\_Echoed\xspace}
\newcommand{\Established}[0]{Established\xspace}
\newcommand{\ShutdownAckSent}[0]{Shutdown\_Ack\_Sent\xspace}
\newcommand{\ShutdownReceived}[0]{Shutdown\_Received\xspace}
\newcommand{\ShutdownSent}[0]{Shutdown\_Sent\xspace}
\newcommand{\ShutdownPending}[0]{Shutdown\_Pending\xspace}
\newcommand{\UserAssoc}[0]{User\_Assoc\xspace}
\newcommand{\UserShutdown}[0]{User\_Shutdown\xspace}
\newcommand{\UserAbort}[0]{User\_Abort\xspace}
\newcommand{\Init}[0]{\texttt{INIT}\xspace}
\newcommand{\InitAck}[0]{\texttt{INIT\_ACK}\xspace}
\newcommand{\CookieError}[0]{\texttt{COOKIE\_ERROR}\xspace}
\newcommand{\CookieEcho}[0]{\texttt{COOKIE\_ECHO}\xspace}
\newcommand{\Data}[0]{\texttt{DATA}\xspace}
\newcommand{\DataAck}[0]{\texttt{DATA\_ACK}\xspace}
\newcommand{\CookieAck}[0]{\texttt{COOKIE\_ACK}\xspace}
\newcommand{\Shutdown}[0]{\texttt{SHUTDOWN}\xspace}
\newcommand{\ShutdownAck}[0]{\texttt{SHUTDOWN\_ACK}\xspace}
\newcommand{\ShutdownComplete}[0]{\texttt{SHUTDOWN\_COMPLETE}\xspace}

\newcommand{\Abort}[0]{\texttt{ABORT}\xspace}
\newcommand{\spin}[0]{\textsc{SPIN}\xspace}
\newcommand{\korg}[0]{\textsc{Korg}\xspace}
\newcommand{\promela}[0]{\textsc{Promela}\xspace}
\newcommand{\amodel}[0]{attacker model\xspace}
\newcommand{\amodels}[0]{attacker models\xspace}
\newcommand{\Amodel}[0]{Attacker Model\xspace}
\newcommand{\Amodels}[0]{Attacker Models\xspace}

\newlist{Qenumerate}{enumerate}{1}
\setlist[Qenumerate]{label={Q}\arabic*}
\newlist{Cenumerate}{enumerate}{1}
\setlist[Cenumerate]{label={C}\arabic*}

\AtBeginDocument{%
  }

\begin{document}

\date{}

\title{A Formal Analysis of SCTP: Attack Synthesis and Patch Verification}

\author{
{\rm Jacob Ginesin}\thanks{Contributed equally.}\\
ginesin.j@northeastern.edu\\
Northeastern University
\and
{\rm Max von Hippel}\footnotemark[1]\\
vonhippel.m@northeastern.edu\\
Northeastern University
\and
{\rm Evan Defloor}\thanks{Listed alphabetically.}\\
defloor.e@northeastern.edu\\
Northeastern University
\and
{\rm Cristina Nita-Rotaru}\footnotemark[2]\\
c.nitarotaru@northeastern.edu\\
Northeastern University
\and
{\rm Michael T{\"u}xen}\footnotemark[2]\\
tuexen@fh-muenster.de\\
FH M{\"u}nster
}

\maketitle


\begin{abstract}
SCTP is a transport protocol 
offering features such as multi-homing, 
multi-streaming, and message-oriented delivery.
Its two main implementations 
were subjected to conformance tests using the \textsc{PacketDrill}
tool. Conformance testing is not exhaustive and 
a recent vulnerability (CVE-2021-3772) showed
SCTP is not immune to attacks.
Changes addressing the vulnerability were implemented,
but the question remains whether other flaws might persist in the protocol design.

We study 
the security of the SCTP design, taking a rigorous approach rooted in formal methods.
We create a formal \promela model of SCTP, 
and define ten properties capturing
the essential protocol functionality based on its RFC specification and 
consultation with the lead RFC author.  Then we show
using the \spin model checker that our model satisfies these properties.
We next define
four representative \amodels \space -- 
Off-Path, where the attacker is an outsider that can spoof the port and IP of a peer;
Evil-Server, where the attacker is a malicious peer;
Replay, where an attacker can capture and replay, but not modify, packets; and
On-Path, where the attacker controls
the channel between peers.
We modify an attack synthesis tool designed for transport protocols, \textsc{Korg},
to support our SCTP model and four \amodels.

We synthesize fourteen unique attacks using the attacker models --
including the vulnerability reported in
CVE-2021-3772 in the Off-Path \amodel,
four attacks in the Evil-Server \amodel,
an opportunistic \Abort attack in the Replay \amodel, and
eight connection manipulation attacks in the On-Path \amodel.  
We show that the proposed patch 
eliminates the vulnerability and does not introduce new ones
according to our model and protocol properties.
Finally, we identify and analyze an ambiguity in the SCTP RFC.
Using the \textsc{AllenNLP} coreference resolution model, we show that the ambiguous text could be 
reasonably interpreted in two ways; then we
model the incorrect interpretation, and synthesize a novel attack against it.
To avoid novice implementers incorrectly interpreting the
RFC, we propose an erratum,
and using the same \textsc{AllenNLP} model,
we show that it 
eliminates the ambiguity.  
\end{abstract}

\section{Introduction}\label{sec:introduction}

Transport protocols play a crucial role in transmitting data across the Internet
either directly -- as in UDP~\cite{rfc768_udp} and DCCP~\cite{rfc4340_dccp}, which provide unreliable communication, and TCP~\cite{rfc9293_tcp_new} and SCTP~\cite{rfc9260}, which provide
reliable communication -- or by supporting secure protocols -- e.g., UDP supports DTLS~\cite{rfc_dtls} and QUIC, 
while TCP supports TLS~\cite{rfc8446_tls13}.
Thus, it is critical that 
transport protocols are designed and implemented to be  bug-free and secure.

SCTP is a transport layer protocol proposed as an alternative to TCP,
offering new features, such as multi-homing, 
multi-streaming, and message-oriented delivery.
Among other use-cases, it is the data channel for WebRTC~\cite{webrtcSCTP}, 
which is used by such applications as Facebook Messenger~\cite{fbWebRTC},
Microsoft Teams~\cite{teamsWebRTC},
and Discord~\cite{discordWebRTC}.
The design of SCTP
is described in RFC documents,
the most recent one being RFC 9260~\cite{rfc9260}, and implemented in Linux~\cite{linux} and FreeBSD~\cite{freebsd}.
These implementations were tested using  \textsc{PacketDrill}~\cite{cardwell2013packetdrill,packetDrillSCTP} and analyzed with \textsc{WireShark}~\cite{rungeler2012sctp}.
Some limited efforts also analyzed the SCTP design using formal methods.
The works in~\cite{vanit2008towards,vanit2013validating} 
focused only on bugs and did not consider attacks,
while the work in~\cite{saini2017evaluating} focused on attacks,
but modeled only 
limited aspects of connection establishment to compare
the resilience of SCTP and TCP to SYN-FLOOD attacks.
A recent vulnerability  -- CVE-2021-3772~\cite{cve} -- shows the importance of conducting a much
more comprehensive 
formal analysis. 
Although a patch was proposed in RFC 9260~\cite{rfc9260}, and adapted by FreeBSD, the question remains whether other flaws might persist in the protocol design and whether the patch might have introduced additional vulnerabilities.
To the best of our knowledge, no prior works formally analyzed the 
entire SCTP connection establishment and teardown routines in a security context.

In this work, we 
take an approach rooted in formal methods to study the security of SCTP.
Our approach is based on \emph{attack synthesis},
where the goal is, given a
program that behaves correctly, and an \amodel, 
to find an attack that can lead the program to behave incorrectly.\footnote{This is totally different from program synthesis, 
where the problem is, given some property, to conjure a program that satisfies it.}
Combined with other formal methods, 
such as model checking,
this approach allows us to precisely study the behaviors of protocols such as SCTP under different threat models.

\textbf{Model Design and Verification.}
We start by creating a finite state machine (FSM) model for the SCTP design 
as specified in RFCs 4960~\cite{rfc4960} and 9260~\cite{rfc9260}, and
writing ten properties the models should satisfy based on a 
close reading of the RFC documents and discussions with the lead SCTP author \footnote{We do not seek to construct a complete set of properties, as we're interested in studying the security-relevant behaviors of SCTP rather than creating an all-encompassing specification. Also, defining a complete specification in LTL is impractical, as LTL is optimized for efficient model checking.}.
Our properties are defined in Linear Temporal Logic (LTL) 
and characterize
the standard establishment and teardown routines,
the proper functioning of the cookie timer,
and the fact that SCTP does not support half-open connections.
Using the \spin model checker, we automatically verify that 
our SCTP model
meets these properties (behaves correctly)
when not under attack.

\textbf{Attack Synthesis.}
Next we define four \amodels (Off-Path, Evil-Server, Replay, and On-Path),  which are representative for transport protocols and provide a wide range of attacker capabilities allowing us to understand the behavior of SCTP when under attack.
The Off-Path \amodel describes 
an attacker who 
may or may not know the IP address or port of either peer,
but cannot read the messages in-transit, and does not know the authentication secrets 
(which in SCTP are called the ``vtags'') of the association.
Thus, its injected messages should theoretically be ignored.
In the Evil-Server \amodel, one peer in an association is malicious, and aims to guide the other peer into some vulnerable state. The Replay \amodel describes an attacker capable of capturing messages from the communication channel and replaying them without modification.
In the On-Path \amodel, the attacker controls the channel connecting the peers, and can intercept, drop, and inject authenticated messages at-will.

We use an attack synthesis
tool for transport protocols called \korg, based on LTL model-checking~\cite{von2020automated}.
We automatically synthesize attacks against our SCTP model,
for each LTL property and \amodel.
In the Off-Path case, we automatically find the attack from CVE-2021-3772.
We find numerous attacks in the Evil-Server and On-Path \amodels, e.g., an Evil-Server attack that establishes a connection with the victim peer and then leaves it stranded, and an On-Path attack that injects messages guiding peers into \ShutdownReceived (an illegal passive/passive teardown).  These results highlight the importance of implementation level defenses against an Evil-Server,
and an end-to-end security model to prevent On-Path attacks.
We also find one Replay attack, highlighting the security-criticality of the transmission sequence number (TSN).

\textbf{Patch Verification.}
We next configure the model to include the patch introduced in RFC 9260~\cite{rfc9260}
and show that the patch fixes the problem, i.e. the property that was violated by
the attack is now met under
the \amodel where that attack was discovered. We further show that in all other \amodels, the same attacks exist with or without the patch, so no new vulnerabilities are introduced by the patch.
\korg is sound and complete,
meaning that (1) if it finds an attack then the attack is real (against the model),
and (2) if any attacks against the model and property exist, of the type \korg
looks for, then given sufficient time and memory, \korg will find one~\cite{von2020automated}.
Since the CVE attack is the kind of attack \korg looks for using the Off-Path \amodel,
the fact that we find the attack when the patch is disabled,
but find no such attacks when it is enabled,
 suffices to prove that the attack does not exist in the patched model.

\textbf{RFC Disambiguation.}
Motivated by the fact that CVE-2021-3772 was caused by a lack of clarity in RFC 4960, we carefully analyze the RFCs for ambiguities.
We identify a portion of RFC 9260
that seems ambiguous to us, and confirm using the 
\textsc{AllenNLP} coreference resolution~\cite{gardner2018allennlp}
natural language processing (NLP) model that
the text can in fact be interpreted in two ways.
We confirm which is correct by consulting
with the lead SCTP RFC author;
then model the incorrect interpretation and, using the \spin~\cite{holzmann1997model} model-checker,
show that it is vulnerable to a potentially serious attack where
the attacker can trick a peer in an association into using the wrong vtag.
We propose an RFC erratum and show using the same NLP approach,
that it unambiguously communicates the correct interpretation.
Finally, we use \textsc{packetDrill}~\cite{cardwell2013packetdrill} 
to confirm that the Linux 
and FreeBSD implementations interpret the ambiguous text correctly.
Note, the FreeBSD implementation was co-authored by the lead SCTP RFC author,
so naturally it interprets the RFC correctly.

\textbf{Contributions.} We summarize our contributions:

\noindent~$\bullet$~\textit{Model:} We model the original SCTP RFC~\cite{rfc4960} 
  using \promela.
  Our model can be configured with or without the CVE patch from RFC 9260~\cite{rfc9260}.
  It is endorsed by the lead SCTP RFC author and faithfully captures the SCTP connection and teardown routines, including the exchange of messages, the user-on-the-loop and its commands, and the handling out-of-the-blue packets. 

\noindent~$\bullet$~\textit{Verification:} We formalize ten novel correctness properties for SCTP in LTL based on a close reading of the RFCs 
and use \spin to prove that our model 
 satisfies all ten when no attacker is present.

\noindent~$\bullet$~\textit{Attack Synthesis:} We introduce four \amodels for SCTP.
  Then we modify \korg to support \emph{packets} and \emph{replay attacks}, and use it to 
  synthesize attacks in the context of each \amodel.
  For Off-Path, we rediscover the CVE before the patch was applied, but not after.
  For Evil-Server, we find four attacks that, depending on implementation details, could leave a victim peer deadlocked or stranded in some liveness cycle, unable to automatically de-associate.
  For Replay, we find one attack that, depending on the security of the TSN, could prevent two peers from establishing a connection.
  We find six similar On-Path attacks where the attacker leads the peers into some illegal state or cycle, violating a property.

\noindent~$\bullet$~\textit{Patch Verification:} 
We show that the patch fixes the problem, i.e. the property that was violated by
the attack is now met under
the \amodel wherein the CVE attack was discovered. Moreover, we show that no new attacks are made possible by the patch in any of the \amodels against any of our ten properties.

\noindent~$\bullet$~\textit{RFC Disambiguation:} 
We identify an ambiguity in RFC 9260 which, 
we show, could be reasonably misinterpreted in a way that 
opens the protocol
 to a new vulnerability.
 We confirm that neither implementation makes the mistake, and 
  to avoid it in future implementations, 
 we suggest an RFC erratum which we show to be unambiguous.

\textbf{Ethics.} We disclosed all of our results to the chair of the SCTP RFC committee.

\textbf{Code.}  All of our results are reproducible with our open source code, available at \url{https://github.com/sctpfm}.

\section{SCTP }\label{sec:background}
In this section we overview SCTP and previous efforts to validate it,
as well as our approach to analyzing its security.

\subsection{Overview}

SCTP is a transport 
protocol proposed as an alternative to TCP,
offering enhanced performance, security features, 
and greater flexibility.
It is specified in several RFCs, each introducing significant modifications. 
RFC 9260~\cite{rfc9260}, which obsolesced RFC 4960~\cite{rfc4960}, 
made numerous small clarifications and improvements, 
including a critical patch for CVE-2021-3772. 
On the other hand, RFC 4960, which obsolesced the original
specification in RFC 2960~\cite{rfc2960}, 
introduced major structural
changes to the protocol as described 
in the errata RFC 4460~\cite{rfc4460}. 
SCTP is implemented in Linux~\cite{linux} and FreeBSD~\cite{freebsd}.

SCTP is a two peer protocol where each peer runs the same state machine.  
However, during connection establishment, the two peers play different roles -- while one peer progresses through the \emph{active} routine in the state machine, the other peer must take a corresponding sequence of \emph{passive} transitions.\footnote{SCTP also supports an initialization routine where both peers are active, called ``initialization collision''.  However, this routine is described in the RFC as an edge-case, rather than an intended use-case.  
}
For teardown there are two options: graceful or graceless.
During graceful tear-down, one peer can act actively and the other passively, or they can both take an active role.
Graceless teardown happens in a single step.

\textbf{Peer States.}
An SCTP peer is identified by a set of IP addresses and a port number.
At any given time, each peer exists in one of finitely many states:
\Closed, in which no association exists;
\CookieWait and \CookieEchoed, used during active establishment;
\Established, in which an association exists and data can be transferred;
\ShutdownReceived and \ShutdownAckSent, used by the passive peer during teardown; and
\ShutdownPending and \ShutdownSent, used by the active peer during teardown.
In active/active teardown, both peers use \ShutdownAckSent.

\textbf{Packets.}
An SCTP packet consists of a common header and a number of chunks.
An essential component of the connection establishment design is authentication of packets
between the peers using a random integer called the \emph{verification tag}, or vtag,
which is initialized using an \emph{initiate tag}, or itag, during establishment.
The
packet header contains the source and destination port number, vtag,
and a checksum. 
The chunk types are \Init, \InitAck, \CookieEcho, and \CookieAck, used during establishment; 
\Data and \DataAck, used for data transmission once an association has been established;
\texttt{ERROR}, used to communicate when an error has occurred;
\Shutdown, \ShutdownAck, and \ShutdownComplete, used during graceful teardown;
\Abort, used during graceless teardown;
and \texttt{HEARTBEAT} and \texttt{HEARTBEAT\_ACK}, used for crash detection.
Chunks contain parameters, e.g., 
 \Init and \InitAck chunks (but no others) contain an itag,
 and (only) \InitAck chunks contain a \emph{state cookie},
 which includes a message authentication code, 
 a timestamp indicating when the cookie was created, and a cookie lifespan.
 There are various kinds of \texttt{ERROR} chunks,
each indicating a different error condition, e.g., \CookieError which indicates receipt of a valid but expired state cookie.
Much like TCP, SCTP uses sequence numbers, called Transmission Sequence Numbers (TSNs).
The initial TSN in an association is proposed by an active participant in the connection
establishment routine, and is incremented with each data transmission thereafter.

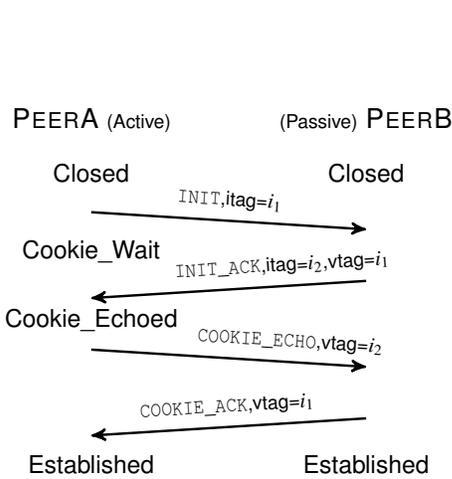
\begin{figure}
\begin{adjustbox}{width=0.35\textwidth,center}
\begin{tikzpicture}[font=\sffamily,>=stealth',thick,
commentl/.style={text width=3cm, align=right},
commentr/.style={commentl, align=left},]
\node[] (active) {\textsc{PeerA} {\scriptsize (Active)}};
\node[right=1cm of active] (passive) {{\scriptsize (Passive)} \textsc{PeerB}};
\node[below=0.1cm of active] {\small \Closed};
\node[below=0.1cm of passive] {\small \Closed};
\node[below=1cm of active] {\small \CookieWait};
\node[below=1.8cm of active] {\small \CookieEchoed};
\draw[->] ([yshift=-.8cm]active.south) 
            coordinate (fin1o) -- 
            ([yshift=-.2cm]fin1o-|passive) 
            coordinate (fin1e) 
            node[pos=.5, above, sloped] {\scriptsize \Init,itag=$i_1$};
\draw[->] ([yshift=-0.8cm]fin1o-|passive) 
            coordinate (fin1e) -- 
            ([yshift=-.2cm]fin1e-|active) 
            coordinate (fin2e) 
            node[pos=.3, above, sloped] {\scriptsize \InitAck,itag=$i_2$,vtag=$i_1$};
\draw[->] ([yshift=-0.8cm]fin1e-|active) 
            coordinate (ceo) -- 
            ([yshift=-.2cm]ceo-|passive) 
            coordinate (cee) 
            node[pos=.72, above, sloped] {\scriptsize \CookieEcho,vtag=$i_2$};
\draw[->] ([yshift=-0.8cm]ceo-|passive) 
            coordinate (cao) -- 
            ([yshift=-.2cm]cao-|active) 
            coordinate (cae) 
            node[pos=.5, above, sloped] {\scriptsize \CookieAck,vtag=$i_1$};

\node[below=0.1cm of cae] {\small \Established};
\node[below=0.9cm of cee] {\small \Established};
\end{tikzpicture}
\end{adjustbox}
\caption{Message sequence chart illustrating SCTP active/passive association establishment routine.  Arrows indicate communication direction and time flows from the top down.}
\label{fig:activePassiveEstablishment}
\end{figure}

\textbf{Connection Establishment.}
In active/passive establishment (Figure~\ref{fig:activePassiveEstablishment}), the active peer sends a packet with
an \Init chunk, containing a nonzero random itag. For the remainder of the association, this (active) peer will only accept packets from the passive peer
that contain a vtag equal to the itag in the \Init it sent.
The passive peer replies with a packet containing an \InitAck
chunk, which also contains a nonzero random itag. 
For the remainder of the association, the passive peer
will only accept packets which contain this itag value as the vtag in the
common header. By checking the vtag, each peer protects itself
from processing packets sent by an attacker not knowing the recipient's vtag.  

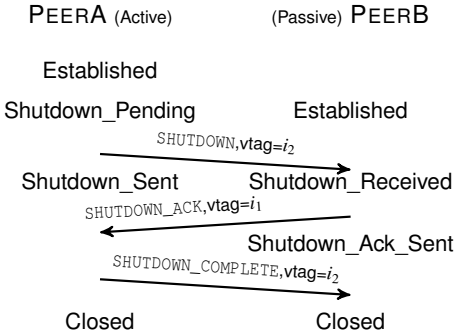
\begin{figure}
\begin{adjustbox}{width=0.35\textwidth,center}
\begin{tikzpicture}[font=\sffamily,>=stealth',thick,
commentl/.style={text width=3cm, align=right},
commentr/.style={commentl, align=left},]
\node[] (active) {\textsc{PeerA} {\scriptsize (Active)}};
\node[right=1cm of active] (passive) { {\scriptsize (Passive)} \textsc{PeerB}};
\node[below=0.2cm of active] {\small \Established};
\node[below=0.7cm of active] {\small \ShutdownPending};
\node[below=0.7cm of passive] {\small \Established};
\node[below=1.6cm of active] {\small \ShutdownSent};
\node[below=1.6cm of passive] {\small \ShutdownReceived};
\node[below=2.4cm of passive] {\small \ShutdownAckSent};
\draw[->] ([yshift=-1.5cm]active.south) 
            coordinate (fin1o) -- 
            ([yshift=-.2cm]fin1o-|passive) 
            coordinate (fin1e) 
            node[pos=.5, above, sloped] {\scriptsize \Shutdown,vtag=$i_2$};
\draw[->] ([yshift=-0.8cm]fin1o-|passive) 
            coordinate (fin1e) -- 
            ([yshift=-.2cm]fin1e-|active) 
            coordinate (fin2e) 
            node[pos=.7, above, sloped] {\scriptsize \ShutdownAck,vtag=$i_1$};
\draw[->] ([yshift=-0.8cm]fin1e-|active) 
            coordinate (ceo) -- 
            ([yshift=-.2cm]ceo-|passive) 
            coordinate (cee) 
            node[pos=.5, above, sloped] {\scriptsize \ShutdownComplete,vtag=$i_2$};
\node[below=0.1cm of cee] {\small \Closed};
\node[below=0.9cm of fin2e] {\small \Closed};
\end{tikzpicture}
\end{adjustbox}
\caption{SCTP active/passive association teardown.}
\label{fig:activePassiveTeardown}
\end{figure}

\textbf{Connection Teardown.}
Teardown can occur gracefully, via the active/passive or active/active routines,
or gracelessly, with an \Abort.
During active/passive teardown (Figure~\ref{fig:activePassiveTeardown}),
the active peer sends a \Shutdown chunk, to which the passive peer responds with \ShutdownAck.
The active peer then sends \ShutdownComplete and both transition to \Closed.
Active/active teardown is also possible, in which the peers exchange, in the following order: \Shutdown, \ShutdownAck, and \ShutdownComplete messages.
The third option is that a peer can gracelessly abort a connection by sending an \Abort chunk.
In this case, both peers immediately transition to Closed.
Once the association is closed, the vtags are forgotten, and when either peer enters a new association, it will randomly choose a new itag (to become its vtag).

\textbf{Timers.}
The SCTP connection routines use three timers: Init, Cookie, and Shutdown.
The goal of the Init Timer is to stop the active peer in an establishment routine from getting stuck
waiting forever for the passive peer to respond to its \Init with an \InitAck.
The goal of the Cookie Timer is similar: it stops that same active peer from getting stuck waiting forever for the passive peer to respond to its \CookieEcho.  
The Shutdown Timer plays a similar role but in the teardown routine, stopping the active peer in teardown from getting stuck waiting for a \ShutdownAck.

\textbf{Out-of-the-Blue Packet Handling.}
In SCTP a message is considered \emph{out-of-the-blue} (OOTB) if the recipient cannot determine to which association the message belongs, i.e., if it has an incorrect vtag, or is an \Init with a zero-valued itag.
Specifically, an OOTB message will be discarded if:
1)  it was not sent from a unicast IP, 
2)  it is an \Abort with an incorrect vtag, 
3) it is an \Init with a zero itag or incorrect vtag\footnote{Per RFC 4960, respond with an \Abort having the vtag of the current association.  But per RFC 9260, discard it.},
4) it is a \CookieEcho, \ShutdownComplete, or \CookieError, 
    and is either unexpected in the current state or has an incorrect vtag, or
5)  it has a zero itag or incorrect vtag.

\textbf{Unexpected Packet Handling.}
A message is \emph{unexpected} if it is not OOTB, but nevertheless, the recipient does not expect it.
SCTP handles unexpected packets as described in Algr.~\ref{alg:pktHandler}.

\begin{algorithm}
\caption{Unexpected Packet Handling}
\label{alg:pktHandler}
\begin{algorithmic}
\REQUIRE Unexpected \textit{msg}
\IF{\textit{msg.chunk} = \Init}
    \IF{\textit{state} = \CookieWait \OR \textit{msg} does not indicate new addresses added}
        \STATE Send \InitAck with \textit{vtag} = \textit{msg.itag}
    \ELSE
        \STATE Discard \textit{msg} and send \Abort with \textit{vtag} = \textit{msg.itag}
    \ENDIF
\ELSIF{\textit{msg.chunk} = \CookieEcho}
    \IF{\textit{msg.timestamp} is expired}
        \STATE Send \CookieError
    \ELSIF{\textit{msg} has fresh parameters}
        \STATE Form a new association
    \ELSE 
        \STATE \texttt{/* initialization collision */}
        \STATE Set \textit{vtag} = \textit{msg.vtag} 
        \STATE \textbf{goto} Established
    \ENDIF
\ELSIF{\textit{msg.chunk} = \ShutdownAck}
    \STATE Send \ShutdownComplete with \textit{vtag} = \textit{msg.vtag}
\ELSE
    \STATE Discard \textit{msg}
\ENDIF
\end{algorithmic}
\end{algorithm}

\textbf{Other Functionality.}  
Other functionalities of SCTP  
include a ``tie-tag'' nonce mechanism used to authenticate a reconnecting peer after a restart;
congestion control\footnote{(based on TCP congestion control)};
fragmentation and reassembly of \Data chunks;
chunk bundling;
 support for the Internet Control Message Protocol (ICMP);
 and multihoming.
 We do not consider this functionality in our analysis, and we refer the
 reader to \cite{rfc9260} for more details.

\subsection{Prior Validation}\label{subsec:sec}
\textbf{Conformance testing.}
The Linux and FreeBSD implementations were tested with \textsc{PacketDrill}~\cite{packetDrillSCTP} and fuzz-tests, suggesting they are crash-free and follow the RFCs.
But this does not necessarily imply the \emph{design} 
in the RFCs behaves correctly in the (a) absense or (b) presense of an attacker.

\textbf{Formal analysis.}
For (a), some prior works formally analyzed SCTP using Colored Petri Net models~\cite{martins2003modelagem,wang2008modeling,vanit2008towards,vanit2013validating} in \textsc{CPNTools}.  This software can check for livelocks (i.e. liveness violations) and deadlocks (stuck states), but it cannot model-check arbitrary logical properties, which seriously limits the use-cases for such models.  One prior work studied (b),
modeling the four-way handshake used by SCTP and comparing it to the three-way handshake used by TCP
in the presence of an attacker, with the Uppaal model-checker~\cite{saini2017evaluating}.
However, the model is closed-source and does not include the teardown routine.  It is unclear whether the model includes OOTB or unexpected packet handling.  We summarize the differences between these prior models and our own in Table~\ref{tab:comparison}.  Finally, the IETF published a security memo for SCTP, but it is not a comprehensive analysis, rather, it simply summarizes prior conversations about security from the SCTP user-group~\cite{sctpSecurityMemo}.

\begin{table*}[]
\centering
\setlength{\tabcolsep}{4pt}
\small
\begin{tabular}{llllllllll}
Work                        & RFC          & Open-Source & Establish & Teardown & OOTB & Unexpected & Livelocks & Deadlocks & Properties \\
Martins et. al.~\cite{martins2003modelagem} & 2960         & N            & Y         & Y        & N    & N          & Y         & Y         & N          \\
Blanchet et. al.~\cite{wang2008modeling} & 2960         & N            & Y         & Y        & N    & N          & Y         & Y         & N          \\
Vanit-Anunchai~\cite{vanit2008towards}     & 4960         & N            & Y         & Y        & Y    & Y          & Y         & Y         & N          \\
Vanit-Anunchai~\cite{vanit2013validating}  & 4960         & N            & Y         & Y        & Y    & Y          & Y         & Y         & N          \\
Saini and Fehnker~\cite{saini2017evaluating}  & 4960         & N            & Y         & N        & N    & N          & Y         & Y         & Y          \\
Ours                        & 4960 \& 9260 & Y            & Y         & Y        & Y    & Y          & Y         & Y         & Y         
\end{tabular}
\caption{Prior formal SCTP analyses versus ours.  RFC column reports modeled version, and open-source column reports whether the model is open-source.  The remaining columns report whether the model includes the establish and teardown routines, OOTB logic, or unexpected packet handling; and if it can be used to check for livelocks or deadlocks, or to verify arbitrary properties. 
}
\label{tab:comparison}
\end{table*}

\textbf{CVE-2021-3772 attack and patch.}
As reported in CVE-2021-3772~\cite{cve}, the prior version of SCTP specified in RFCs 2960~\cite{rfc2960} and 4960~\cite{rfc4960}
is vulnerable to a denial-of-service attack.
The reported vulnerability worked as follows.  Suppose SCTP peers A and B have established a connection and an off-channel attacker knows the IP addresses and ports of the two peers, but not the vtags of their existing connection.  The attacker spoofs B and sends a packet containing an \Init
to A.  The attacker uses a zero vtag as required for
packets containing an \Init.  The attacker must use
an illegal parameter in the \Init, e.g., a zero itag.  

\begin{figure}
\begin{adjustbox}{width=0.35\textwidth,center}
\begin{tikzpicture}[font=\sffamily,>=stealth',thick,
commentl/.style={text width=3cm, align=right},
commentr/.style={commentl, align=left},]
\node[] (init) {\textsc{Attacker}};
\node[right=1cm of init] (recv) {\small \textsc{PeerA}};
\node[right=1cm of recv] (third) {\small \textsc{PeerB}};
\node[below=0.1cm of recv] {\small \Established};
\node[below=0.1cm of third] {\small \Established};
\draw[->] ([yshift=-0.8cm]init.south) 
            coordinate (fin1o) -- 
            ([yshift=-.2cm]fin1o-|recv) 
            coordinate (fin1e) 
            node[pos=.3, above, sloped] {\scriptsize \Init,vtag=0,itag=0};
\draw[->] ([yshift=-0.8cm]fin1o-|recv) 
            coordinate (fin1e) -- 
            ([yshift=-.2cm]fin1e-|third) 
            coordinate (fin2e) 
            node[pos=.3, above, sloped] {\scriptsize \Abort,vtag=$i_2$};
\node[below=1.8cm of recv] {\small \Closed};
\node[below=2cm of third] {\small \Closed};
\end{tikzpicture}
\end{adjustbox}
\caption{Attack disclosed in CVE-2021-3772.  Peers A and B begin having  established an association with vtags~$i_1, i_2$ (resp.).  The Attacker transmits an invalid \Init chunk to  A, spoofing the port and IP of B.  Peer A responds by sending a valid \Abort to B, which closes the association.  By sending a single invalid \Init the Attacker performs a DoS.}
\label{fig:cve-msc}
\end{figure}
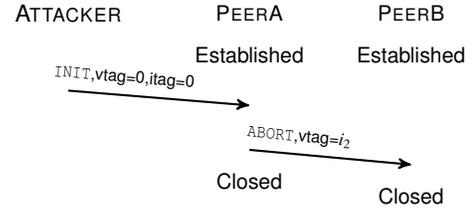

Peer A, having already established a connection, treats the packet as
out-of-the-blue, per RFC 2960 $\mathsection$8.4 and 5.1, which specify that as an
association was established, A should respond to the \Init
containing illegal parameters with an \Abort and go
to \Closed. But in RFCs 2960 and 4960, it is unspecified which vtag should be
used in the \Abort.  Some implementations
used the expected vtag, which is
where a vulnerability arises.
Since the attacker spoofed the IP and port of Peer B,
Peer A sends the \Abort to Peer B, not the attacker.
When Peer B receives the \Abort, it sees the correct vtag, and 
tears down the connection.  Thus, by injecting a single packet with zero-valued tags, the attacker tears down the 
connection, pulling off a DoS.  
The attack is illustrated in Figure~\ref{fig:cve-msc}.

RFC 9260 patches CVE-2021-3772 using a strict defensive measure, wherein OOTB \Init packets with empty or zero itags are discarded, without response.  FreeBSD~\cite{freebsd} uses this patch.
Linux, on the other hand, adopts a different patch~\cite{linuxNotes}, wherein
the peer receiving the \Abort with the zero vtag simply ignores
it (rather than close the connection).

\section{Our SCTP Model}\label{sec:model}
In this section, we describe our SCTP \promela model and properties that guide
our analysis.

\subsection{Overview}%
\label{sub:Overview}
As we are primarily interested in denial-of-service attacks, 
and in order to avoid state-space explosion,
we selectively model the SCTP connection establishment and teardown routines.
This allows us to automatically and exhaustively explore, simulate, and verify the critical, security-relevant aspects of SCTP. 
Our model captures the following aspects of SCTP per RFC 9260~\cite{rfc9260}: internal peer states, packet verification using the itag and vtag, timers, TSNs, and handling for invalid, unexpected, and OOTB packets. 
We made only the abstractions listed in Section~\ref{subsec:abstractions}.

Although our model is fully faithful to the SCTP RFC~\cite{rfc9260}, and is an executable program, it is not a network library and cannot be used in place of the existing Linux or FreeBSD implementations.  This is because of the abstractions and simplifications mentioned above, and also because it does not implement API hooks for higher-level applications, nor syscalls to  transmit over the Internet.  It is simply a model of SCTP with which we can formally verify correctness properties.

\subsection{Model Details}%
\label{sub:Model Structure}

We describe our SCTP model in \promela, focusing on internal peer states, packet verification, invalid packet defense mechanisms, timeouts, and OOTB packet handling.

\noindent~\textbf{Mathematical Preliminaries.}
Linear Temporal Logic is a modal logic for reasoning about program executions.
In LTL, we say a program $P$ \emph{models} a property $\phi$,
written $P \models \phi$, if $\phi$ holds over every execution of $P$.
If $\phi$ holds over some but not all executions of $P$, then we write $P \centernot{\models} \phi$.

The LTL language is given by predicates (e.g., ``Peer A is in \Established'' or ``Peer B's cookie timer is inactive''); the temporal operators ``next'', ``always'', ``eventually'', and ``until''; and the logical operators of negation, conjunction, and disjunction.
An LTL model-checker is a 
tool that, given $P$ and $\phi$, can automatically check whether or not $P \models \phi$ \footnote{LTL model-checking is decidable, and reduces to checking B\"uchi Automata intersection emptiness, which is PSPACE-complete.}. We use the model-checker \spin\footnote{version 6.5.2}, whose language is \promela.

We use $\parallel$ to denote rendezvous 
composition, so,
$S = P \parallel Q$ denotes that the program $S$ equals the composition of $P$ with $Q$.
Specifically, matching \emph{send} transitions of $P$ and \emph{receive} transitions of $Q$ occur in lockstep, and vice versa.  Note, in our model, we actually build a process called a ``channel'' to capture network delay, and we rendezvous-compose the channel with the two peers to build asynchronous communication (which is more realistic).
The $\parallel$ operator is commutative and associative.
For more details, refer to $\mathsection$2 of~\cite{von2020automated}.

\noindent~\textbf{Internal Peer States.}
Our model consists of two peers (A and B) and a channel connecting them. That is, we study the system $S = \textsc{PeerA} \parallel \textsc{Channel} \parallel \textsc{PeerB}$.  Each peer is represented by an identical FSM, illustrated in Fig.~\ref{fig:fsm}. Transitions between states occur based on the receipt of user commands, or communication and message processing.

The channel connecting the two peers contains an internal single-message buffer in each direction (meaning it can hold two messages at once, one traveling from left to right and the other from right to left).  It does not drop, corrupt, nor create messages, and cannot accept a new message in a given direction until the old one was delivered.  In other words, it is lossless and FIFO, in that it guarantees every delivered message was sent and messages are delivered in order.  The entire setup is illustrated in Figure~\ref{fig:network}.

\begin{figure}
\begin{adjustbox}{width=0.45\textwidth,center}
\begin{tikzpicture}
\node[draw, rectangle, rounded corners, minimum width=1.7cm, minimum height=1.5cm, fill=white] (channel) at (3,0) 
    {\small \textsc{Channel}};
\node[draw,rectangle] (AtoB) at (3,-0.4) {\small \textsc{AtoB}};
\node[draw,rectangle] (BtoA) at (3,0.4) {\small \textsc{BtoA}};
\draw[->] (0.5,-0.4) to (AtoB);
\draw[->] (AtoB) to (4.49,-0.4);
\draw[->] (5,0.4) to (BtoA);
\draw[->] (BtoA) to (1.49,0.4);
\node[draw, circle, minimum height=1.2cm,fill=white] (sender) at (1,0) 
    {\small \textsc{PeerA}};
\node[draw, circle, minimum height=1.2cm,fill=white] (receiver) at (5,0) 
    {\small \textsc{PeerB}};
\node[draw,rectangle,rounded corners,left=0.4cm of sender,align=center] (userA) {\small \textsc{UserA}};
\node[draw,rectangle,rounded corners,right=0.4cm of receiver,align=center] (userB) {\small \textsc{UserB}};
\draw[<->] (userA) to (sender);
\draw[<->] (userB) to (receiver);
\end{tikzpicture}
\end{adjustbox}
\caption{The system $\textsc{UserA} \parallel \textsc{PeerA} \parallel \textsc{Channel} \parallel \textsc{PeerB} \parallel \textsc{UserB}$. \textsc{Channel} contains a size-1 FIFO buffer in each direction (AtoB and BtoA, respectively).  Arrows indicate communication direction.  Composition between \textsc{Channel} and peers is rendezvous (the buffers are inside \textsc{Channel}).}
\label{fig:network}
\end{figure}
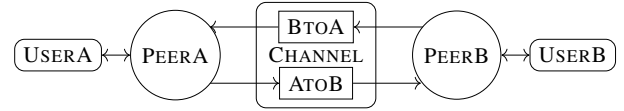

\begin{figure*}
\begin{adjustbox}{width=0.89\textwidth,center}
\begin{tikzpicture}
\begin{scope}[every node/.style={rectangle,draw,minimum width=2em,minimum height=2em}]
    \node (CLOSED) at (1.75,5.36) {\scriptsize \Closed}; 
    \node (COOKIEWAIT) at (-1.5,3.3) {\scriptsize \CookieWait};
    \node (COOKIEECHOED) at (-1.5,1) {\scriptsize \CookieEchoed};
    \node (ESTABLISHED) at (1.75,-0.4) {\scriptsize \Established};
    \node (SHUTDOWNRCVD) at (6,1) {\scriptsize \ShutdownReceived};
    \node (SHUTDOWNPENDING) at (9,1) {\scriptsize \ShutdownPending};
    \node (SHUTDOWNSENT) at (9,3.3) {\scriptsize \ShutdownSent};
    \node (SHUTDOWNACKSENT) at (6,3.3) {\scriptsize \ShutdownAckSent};

\begin{scope}[>={Stealth[black]},
              every node/.style={fill=white,rectangle},
              every edge/.style={draw=black,thick}]
    \path [->] (CLOSED) edge [loop above] 
              node[yshift=0.05cm] {\scriptsize \Init,\texttt{N},\texttt{E}? \InitAck,\texttt{E},\texttt{E}!} 
              (CLOSED);
    \path [->] (-0.5,5.36) edge 
              (CLOSED);
    \path [->] (CLOSED) edge
              node[left,xshift=-0.4cm] {\scriptsize \UserAssoc? \Init,\texttt{N},\texttt{E}!}
              (COOKIEWAIT);
    \path [->] (CLOSED) edge
              node[text width=1cm,yshift=-0.3cm,xshift=0.7cm,fill=none] 
              {\scriptsize \CookieEcho,\texttt{E},\texttt{N}?\\\CookieAck,\texttt{E},\texttt{N}!}
              (ESTABLISHED);
    \path [->] (COOKIEWAIT) edge 
              node[below right,text width=2cm,yshift=0.4cm,fill=none] 
              {\scriptsize \InitAck,\texttt{E},\texttt{E}?\\\CookieEcho,\texttt{E},\texttt{N}!} 
              (COOKIEECHOED);
    \path [->] (COOKIEECHOED) edge 
              node[yshift=0.1cm,xshift=-0.5cm,text width=1.1cm] {\scriptsize \CookieAck,\texttt{E},\texttt{N}?} 
              (ESTABLISHED);
    \path [->] (COOKIEECHOED) edge [loop left]
               node[below,yshift=-0.35cm,xshift=0.1cm] 
               {\scriptsize \CookieError,\texttt{E},\texttt{N}?}
               node[below,yshift=-0.7cm,xshift=0.1cm] 
               {\scriptsize \textit{then optionally,} \Init,\texttt{N},\texttt{E}!}
               (COOKIEECHOED);
    \path [->] (COOKIEECHOED.north west) edge [bend left]
               node[above left,xshift=-0.1cm] {\scriptsize\CookieError,\texttt{E},\texttt{N}?}
               node[below left,xshift=-0.1cm] {\scriptsize \Init,\texttt{N},\texttt{E}!}
               (COOKIEWAIT.south west);
    \path [->] (ESTABLISHED) edge [out=east,in=south,looseness=0.5] 
              node[xshift=-1cm,yshift=0.2cm,fill=none] {\scriptsize \UserShutdown?} 
              (SHUTDOWNPENDING);
    \path [->] (ESTABLISHED) edge 
              node[xshift=1.1cm,fill=none] {\scriptsize \Shutdown,\texttt{E},\texttt{N}?} 
              (SHUTDOWNRCVD.south west);
    \path [->] (SHUTDOWNPENDING) edge 
              node[yshift=-0.2cm,xshift=1cm] {\scriptsize \Shutdown,\texttt{E},\texttt{N}!} 
              (SHUTDOWNSENT);
    \path [->] (SHUTDOWNSENT) edge [bend right] 
              node[xshift=2.5cm,text width=3cm,fill=none] 
              {\scriptsize \ShutdownAck,\texttt{E},\texttt{N}?\\\ShutdownComplete,\texttt{E},\texttt{N}!}
              (CLOSED);
    \path [->] (SHUTDOWNSENT) edge 
              node[below,xshift=0.1cm,yshift=-0.4cm] {\scriptsize \Shutdown,\texttt{E},\texttt{N}?} 
              node[below,yshift=-0.7cm] {\scriptsize \ShutdownAck,\texttt{E},\texttt{N}!} 
              (SHUTDOWNACKSENT);
    \path [->] (SHUTDOWNRCVD) edge 
              node[below] {\scriptsize \ShutdownAck,\texttt{E},\texttt{N}!} 
              (SHUTDOWNACKSENT);
    \path [->] (SHUTDOWNACKSENT) edge [bend right] 
              node[above,xshift=-0.9cm,yshift=-1.5cm,text width=2cm,fill=none] 
              {\scriptsize \texttt{SHUTDOWN\_}\\\texttt{COMPLETE},\texttt{E},\texttt{N}?\\\textit{or} (\ShutdownAck,\texttt{E},\texttt{N}?\\\ShutdownComplete,\texttt{E},\texttt{N}!)} 
              (CLOSED);
\end{scope}
\end{scope}

\begin{scope}[on background layer]
    \fill[white] (0,0) rectangle (5,5);
  \end{scope}
\end{tikzpicture}
\end{adjustbox}
\caption{SCTP Finite State Machine.  $x,v,i?$ (or $x,v,i!$) denotes receive (or send) chunk~$x$ with vtag~$v$ and itag~$i$.  Events in multi-event transitions occur in the order they are listed.  Logic for OOTB packets, \Abort messages or \UserAbort commands, unexpected user commands, and data exchange are ommitted but faithfully implemented in the model and described in this paper.}
\label{fig:fsm}
\end{figure*}


\noindent~\textbf{Packet Verification and Invalid Packet Defenses.}
We model each SCTP message as consisting of a message chunk, 
a vtag, and an itag.
Each of these components are modeled using enums, which in \textsc{Promela} are called \texttt{mtype}s.
The message chunk denotes the meaning of the message, e.g., a message with an \Init chunk is called an \emph{initiate message} and is used to initiate a connection establishment routine. 
The itag and vtag are used to verify the authenticity of the sender of the message, as described in Section~\ref{subsec:sec}.  In our model there are three kinds of tags: expected (\texttt{E}), unexpected (\texttt{U}), or none (\texttt{N}).  A tag is expected if (1) it is a non-zero itag on an \Init or \InitAck chunk, or (2) it is the other peer's vtag in the existing association.  Otherwise, it is unexpected.  The none type is reserved for packets that do not carry the given tag type -- e.g., only \Init and \InitAck chunks carry an itag, so in the other types of messages, the itag is \texttt{N}. The BNF grammar for messages in our model is given in Figure~\ref{fig:bnf}.
\begin{figure}
\[
\begin{aligned}
\textit{msg} & ::= \Init,\texttt{N},\textit{ex} \mid
                 \InitAck,\textit{ex},\textit{ex} \mid
                 \textit{ach},\textit{ex},\texttt{N} \\
\textit{ach} & ::= \Abort \mid \Shutdown \mid \ShutdownComplete \\
             & \mid \CookieEcho \mid \CookieAck \mid \ShutdownAck \\
             & \mid \CookieError \mid \Data \mid \DataAck \\
\textit{ex} & ::= \texttt{E} \mid \texttt{U}
\end{aligned}
\]
\caption{BNF grammar for messages in our model.}
\label{fig:bnf}
\end{figure}
We also support an option where the \textit{msg} can be extended with a TSN.

Upon receiving a message, our model checks that the tags are set as expected,
depending on the message and state.
If a message has an unexpected tag then
the model employs the defenses specified in the RFC, e.g., silently discarding the message or responding with an \Abort. 
These 
defenses can be configured with or without the CVE patch from RFC 9260.

\noindent~\textbf{Active and Passive Connection Routines.}
Our SCTP model implements active/passive establishment and teardown, as well as active/active teardown, but not active/active establishment (a.k.a. ``\Init collision''), precisely as described in Section~\ref{sec:background} and illustrated in Figures~\ref{fig:activePassiveEstablishment} and~\ref{fig:activePassiveTeardown}, with the caveat that the itag and vtag are abstracted (as described above).
We also capture the TSN proposal and use throughout an association, although this feature can be
turned off in our model to reduce the state-space
for more efficient verification.

\noindent~\textbf{Out-of-the-Blue and Unexpected Packets.}  Our model faithfully captures OOTB logic described in $\mathsection$8.4 of RFCs 4960 and 9260, with only the exceptions given in Section~\ref{subsec:abstractions}.

\subsection{Ambiguity in the RFC}\label{subsec:ambiguity}
We found one ambiguity in the SCTP RFCs, in $\mathsection$5.2.1, during the description of how a peer should react upon receiving an unexpected \Init chunk:
\begin{myquote}
\textit{Upon receipt of an \Init chunk in the \CookieEchoed state, an endpoint
   MUST respond with an \InitAck chunk using the same parameters it sent
   in its original \Init chunk (including its Initiate Tag, unchanged),
   provided that no new address has been added to the forming
   association.}
\end{myquote}
Consider two peers - A and B - initially both in \Closed, in addition to some attacker who can spoof the port and IP  of B.  Suppose these machines engage in the sequence of events illustrated in Figure~\ref{fig:ambiguity}.  At the end of the sequence, what value should the vtag $V$ take? 

\begin{figure}
\begin{adjustbox}{width=0.4\textwidth,center}
\begin{tikzpicture}[font=\sffamily,>=stealth',thick,
commentl/.style={text width=3cm, align=right},
commentr/.style={commentl, align=left},]
\node[] (active) {\textsc{PeerA} {\scriptsize (Active)}};
\node[right=1cm of active] (passive) {{\scriptsize (Passive)} \textsc{PeerB}};
\node[right=0.5cm of passive] (attacker) {\textsc{Attacker}};
\node[below=0.1cm of active] {\small \Closed};
\node[below=0.1cm of passive] {\small \Closed};
\node[below=1cm of active] {\small \CookieWait};
\node[below=1.8cm of active] {\small \CookieEchoed};
\draw[->] ([yshift=-.8cm]active.south) 
            coordinate (fin1o) -- 
            ([yshift=-.2cm]fin1o-|passive) 
            coordinate (fin1e) 
            node[pos=.5, above, sloped] {\scriptsize \Init,itag=$i_1$};
\draw[->] ([yshift=-0.8cm]fin1o-|passive) 
            coordinate (fin1e) -- 
            ([yshift=-.2cm]fin1e-|active) 
            coordinate (fin2e) 
            node[pos=.3, above, sloped] {\scriptsize \InitAck,vtag=$i_1$,itag=$i_2$};
\draw[->] ([yshift=-1cm]fin1e-|attacker)
            coordinate (attack1f) --
            ([yshift=-1cm]fin2e-|active)
            coordinate (fin2f)
            node[pos=.3, above, sloped] {\scriptsize \Init,itag=$i_3$};
\draw[->] ([yshift=-.2cm]fin2f-|active)
            coordinate (response1g) --
            ([yshift=-.8cm]attack1f-|fin1e)
            coordinate (fin2g)
            node[pos=.8, above, sloped] {\scriptsize \InitAck,vtag=$V$,itag=$i_1$};
\end{tikzpicture}
\end{adjustbox}
\caption{Ambiguous scenario.  What value should $V$ take?  See Section~\ref{subsec:ambiguity}.}
\label{fig:ambiguity}
\end{figure}

To understand all possible interpretations, we ran the text through the 
\textsc{AllenNLP} Coreference Resolution tool~\cite{gardner2018allennlp},
a state-of-the-art NLP model trained to detect which words in a sentence refer to the same entity, producing the following output, where entities are given 
ids and colored for readability; and each reference~$R$ to an entity with id~$I$ is highlighted \bluebox{\lightbluebox{$I$} $R$}.

\begin{leftbar}
Upon receipt of an \pinkbox{\lightpinkbox{1} \Init} chunk in the \CookieEchoed state, 
\bluebox{\lightbluebox{0} an endpoint}
   MUST respond with an \InitAck chunk using the same parameters 
   \bluebox{\lightbluebox{0} it}
   sent in 
   \bluebox{\lightbluebox{0} its}
   original \pinkbox{\lightpinkbox{1} \Init} chunk 
   (including \bluebox{\lightbluebox{0} its} Initiate Tag, unchanged),
   provided that no new address has been added to the forming
   association.
\end{leftbar}

The model predicts that the occurrences of \emph{it} and \emph{its} all refer to the same entity as \emph{an endpoint}, which is clearly the responding endpoint, i.e., Peer A.
What if a reader interprets ``the same parameters'' to include the vtag?
Then the model would predict that the vtag of the \InitAck should come from the \Init that entity 
\bluebox{\lightbluebox{0}} sent, implying $V$ should take the itag of the message, i.e. $V=i_1$.
The fact that this is wrong only becomes clear if you fully understand how itags and vtags are used in both directions.
To make the text unambiguous, we suggest adding the following sentence:
\begin{myquote}
The verification tag used in the packet containing the \InitAck chunk MUST
be the initiate tag of the newly received \Init chunk.
\end{myquote}
The coreference resolution model predicts that ``the newly received \Init chunk'' is the same entity as the \Init chunk in ``Upon receipt of an \Init chunk in the \CookieEchoed state'', so the text is unambiguous.

\subsection{Abstractions and Limitations}\label{subsec:abstractions}
Our model is fully faithful to the SCTP RFC, modulo the following abstractions and limitations.

\noindent \textbf{$\bullet$~Unicast peers.}  In the RFC, OOTB messages from non-unicast peers are  discarded; we model all peers as unicast.

\noindent \textbf{$\bullet$~No crashes or restarts.}  In our model, peers never crash or restart.  Thus we also ommit crash detection (including \texttt{HEARTBEAT} and \texttt{HEARTBEAT\_ACK} chunks).
   
\noindent \textbf{$\bullet$~Tags are abstracted.}  We do not model tie-tags, which are used when reconnecting a peer to an existing association after a restart.  In the RFC, itags and vtags are integer-valued and chosen randomly.  But we model tags as the ``expected'' value, an ``unexpected'' value, or ``none'', since this level of detail is all that matters for our properties.  A side-effect is that we cannot study \Init collision.  \Init collision is not included in the State Association Diagram in RFC 9260 $\mathsection$4, nor in the various association flows throughout the RFC document, leading us to believe it is not a protocol feature but rather an edge-case the protocol is designed to withstand.

\noindent \textbf{$\bullet$~Perfect channel.} We do not model packet loss, reordering, nor corruption, nor how SCTP deals with these scenarios.
 
\noindent \textbf{$\bullet$~Peers do not exchange data while in \Established.} Because we focus on denial-of-service attacks, modeling data exchange while in \Established is unnecessary; rather, we focused on the connection and disconnection of peers.  We did model data transmission outside of \Established, in case it caused edge-case behaviors during teardown.

\noindent \textbf{$\bullet$~Packets only ever contain one chunk.}  Since we also do not model (or write properties about) fragmentation, bundling, or reassembly, we can simulate multi-chunk transmissions by sending consecutive single-chunk messages.
 
\noindent \textbf{$\bullet$~Simplified packet structure.}  We choose not to model packet structure details relevant to only \Data packets, e.g.: stream sequence number, payload protocol identifier, and variable length.  We also do not model ICMP messages.

\subsection{Correctness Properties}\label{subsec:properties}

Next we transcribe ten logical properties we believe SCTP should satisfy. Note, we do not intend to create a \textit{complete} set of properties that captures all behaviors of SCTP. Rather, we design our properties to capture the security-relevant behavior of SCTP. 
Each property is implemented in \textsc{Promela} using LTL.  We justify each using the RFCs~\cite{rfc4960,rfc9260} and our intuition about the security SCTP should provide.
\begin{description}[style=unboxed,leftmargin=0cm,itemsep=0pt,parsep=\parskip]

\item \textbf{$\phi_1$: A peer in \Closed either stays still or transitions to \Established or \CookieWait.}
This is based on the routine described in $\mathsection$5.1, as well as the Association State Diagram in $\mathsection$4.  If a closed peer could transition to any state other than \Established or \CookieWait, it could de-synchronize with the other peer, breaking the four-way handshake and potentially leading to a deadlock, livelock, or other problem.
\item \textbf{$\phi_2$: One of the following always eventually happens: the peers are both in \Closed, the peers are both in \Established, or one of the peers changes state.}
The property we want to capture here, ``no half-open connections'', is stated in $\mathsection$1.5.1,
was verified in the related work by Saini and Fehnker~\cite{saini2017evaluating},
and was studied for TCP in two prior works~\cite{von2020automated,pacheco2022automated}.
But we have to formalize it subtly, because 
in the case of an in-transit \Abort,
it is possible for one peer to temporarily be in \Established while the other is in \Closed; so we write it as a liveness property, saying half-open states eventually end.
\item \textbf{$\phi_3$: If a peer transitions out of \ShutdownAckSent then it must transition into \Closed.}
We derived this from the Association State Diagram in $\mathsection$4.  Every transition out of \ShutdownAckSent described in the RFC ends up in either \Closed or \ShutdownAckSent.  If this property fails, it would imply a flaw in the graceful teardown routine, and could cause a deadlock, livelock, or other problem.
\item \textbf{$\phi_4$: If a peer is in \CookieEchoed then its cookie timer is actively ticking.}
Per $\mathsection$5.1 C), the peer starts the cookie timer upon entering \CookieEchoed.  Per $\mathsection$4 step 3), when the timer expires it is reset, up to a fixed number of times, at which point the peer returns to \Closed.  
If the property fails, then the active peer in an establishment could get stuck in \CookieEchoed forever,
opening a new opportunity for DoS.
\item \textbf{$\phi_5$: The peers are never both in \ShutdownReceived.}
This property follows from inspection of the Association State Diagram in $\mathsection$4.
From a security perspective, if both peers were in \ShutdownReceived, this would indicate that neither initiated the shutdown (yet both are shutting down); the only logical explanation for which is some kind of DoS.
\item \textbf{$\phi_6$: If a peer transitions out of \ShutdownReceived then it must transition into either \ShutdownAckSent or \Closed.}
The transition to \ShutdownAckSent is shown in the Association State Diagram in $\mathsection$4.
The transition to \Closed can occur upon receiving either a \UserAbort from the user or an \Abort from the other peer.  No other transitions out of \ShutdownReceived are given in the RFC.  If this property fails, it could de-synchronize the teardown handshake, potentially leading to an unsafe behavior.  For example, if a peer transitioned from \ShutdownReceived to \Established, it would end up in a half-open connection.
\item \textbf{$\phi_7$: If Peer~A is in \CookieEchoed then B must not be in \ShutdownReceived.}
We derived this from the Association Diagram in $\mathsection$4, which shows A must receive an \InitAck while in the \CookieWait and then send a \CookieEcho in order to transition into \CookieEchoed. B must have been in \Closed to send an \InitAck in the first place, hence B cannot be in \ShutdownReceived.  This property relates to the synchronization between the peers: if one is establishing a connection while the other is tearing down, then they are de-synchronized, and the protocol has failed.
\item \textbf{$\phi_8$: Suppose that in the last time-step, Peer~A was in \Closed and Peer~B was in \Established.  Suppose neither user issued a \UserAbort, and neither peer had a timer time out.  Then if Peer A changed state, it must have changed to either \Established, or the implicit, intermediary state in \CookieWait in which it received \InitAck but did not yet transmit \CookieEcho.}
The transitions from \Closed to \Established and the described intermediary state are implicit in the Association State Diagram in $\mathsection$4.  The timer caveat is described in $\mathsection$4 step 2, and the aborting caveat is  in $\mathsection$9.1.  If the property fails, the four-way handshake ended, yet was not completed successfully, did not time out, and was not aborted, so somehow, the protocol failed.
\item \textbf{$\phi_9$: The same as $\phi_8$ but the roles are reversed.}
The property is: \emph{Suppose that in the last time-step, Peer~B was in \Closed and Peer~A was in \Established ...}

\item \textbf{$\phi_{10}$: Once connection termination initiates, both peers eventually reach \Closed}. This follows from the description of connection termination in $\mathsection$9. Once connection termination is initiated, there is no way to recover the association.

\end{description} 
For the On-Path \amodel, $\phi_8$ and $\phi_9$ are symmetric.  For the other \amodels, the properties are distinct, because the \amodel's network topology is asymmetric.

\subsection{Validating Our Model}%
\label{sub:Validating Our Model}

Our model allows us to execute and reason about any component of the peer logic in isolation, or two interacting peers.
To verify our model, we extracted the properties listed above from the SCTP RFCs, and then used the model-checker to prove that our model satisfies all of the properties.  We interactively guided \spin to drive the model through various connection flows (which we compared to the RFC text), and we manually compared our logic for handling OOTB packets to the corresponding \textsc{C} code in Linux and FreeBSD.  Finally, we used \spin to prove there were no deadlocks or livelocks (liveness cycles) and all the peer states are reachable.

\section{SCTP Attack Synthesis}\label{sec:synthesis}

In this section we provide details on attack synthesis
and \korg, the tool we used. Next, we describe four attacker 
models we defined and used for our analysis.
Finally, we present the changes we had
to make to \korg to handle our SCTP model, and
the four attacker models we considered.

\subsection{Attack Synthesis}

\emph{LTL program synthesis} is the problem of,
given an LTL specification~$\phi$, 
automatically deriving a compliant program~$P$ (for which $P \models \phi$).
\emph{LTL attack synthesis} is fundamentally different (logically dual),
and cannot be solved using program synthesis alone.
In attack synthesis, the problem is flipped:
given a program $S$ and property $\phi$,
where $S$ is already compliant ($S \models \phi$), 
if $S = P \parallel Q$ consists of an \emph{invariant component} $P$ (that the attacker cannot change) and a \emph{variant component} $Q$ (that the attacker can change), 
we ask whether there exists some modification $A$ such that,
if we replace $Q$ with $A$,
the new system $S' = P \parallel A$ is non-compliant ($S' \centernot{\models} \phi$).
In other words, we study a system that behaves correctly,
and ask if we can change some constrained aspect
so that it behaves incorrectly.
If so, we call this modification~$A$ an ``attack''.

There are multiple kinds of attacks one might try to synthesize,
depending on the nature of the protocol and the attacker goal.
We use use \korg~\cite{von2020automated}, which leverages \spin~\cite{holzmann1997model} to synthesize attacks against arbitrary LTL properties of transport protocols.
\korg was previously successfully applied to TCP and DCCP~\cite{von2020automated,pacheco2022automated}, and
to the best of our knowledge, it is the only open-source attack
synthesis tool that can synthesize terminating fixed-vocabulary communication attacks against
arbitrary LTL properties.\footnote{We discuss another approach~\cite{matsui2022synthesis} based on reactive controller synthesis in Sec.~\ref{sec:related}, but it is not directly comparable as it synthesizes a narrow category of attacks that are guaranteed to succeed, which do not always exist.}
\korg is proven to be sound (it has no false-positives)
and complete (if attacks of the kind \korg looks for exist,
given enough resources, \korg will find one).
Meanwhile, \spin has existed for 35 years; has
been applied to dozens of real systems including the Mars rover~\cite{mars},
PathStar access server~\cite{holzmann2000automating},
and
ISO/IEEE P11073-20601 medical communication protocol~\cite{goga2009formal};
spawned a dedicated formal methods conference, currently in its 30th year\footnote{\url{https://spin-web.github.io/SPIN2023/}}; and earned the the 2002 ACM Software System Award.

On the other hand, 
\textsc{ProVerif}~\cite{blanchet2016modeling} and \textsc{Tamarin}~\cite{meier2013tamarin}
are designed to verify, or synthesize attacks against,
secrecy, authentication, privacy, and equivalence properties of cryptographic protocols
using the Dolev-Yao \amodel~\cite{dolev1983security}.
Although \textsc{Tamarin} admits a small language of guarded first-order trace properties,
in practice it does not scale 
to complicated properties (e.g., $\phi_8$) 
of complex models (e.g., ours)~\cite{ben2014generic}, 
the language is not as well-suited as LTL for verifying state-based properties of transport protocols,
and it does not support arbitrary \amodels (such as ours, which differ from Dolev-Yao -- e.g., because in Dolev-Yao the attacker can spawn infinitely many associations at once, but in our \amodels, it is limited to one at a time).
Thus, \textsc{ProVerif} and \textsc{Tamarin}~\cite{meier2013tamarin} are better
than \korg for studying cryptographic protocols 
such as TLS~\cite{bhargavan2017verified} or Signal~\cite{kobeissi2017automated}
under Dolev-Yao,
but \korg is better for synthesizing communication attacks against
non-cryptographic
transport protocols such as SCTP under custom \amodels.

\korg requires four inputs: 
an invariant component~$P$ (e.g., the SCTP model) and 
variant component~$Q$ (which in our case is part of the \amodel), both in \promela;
an LTL correctness property~$\phi$, such that the composite system consisting
of both $P$ and $Q$ satisfies $\phi$ ($P \parallel Q \models \phi$);
and a YAML file encoding the grammar (I/O) of $Q$
(which become the I/O of the attacker).
\korg generates an model with these inputs in which $Q$ is replaced with a process called a \emph{daisy}, that can nondeterministically send or receive messages specified in the grammar.
Next, it modifies $\phi$ to have a precondition saying the daisy terminates, and then asks \spin to verify or disprove the modified property for the modified model.
Either \korg reports no attacks exist, or \spin outputs a counterexample execution, which \korg parses into an attack~$A$.
For more refer to~\cite{von2020automated}.
The inputs to \korg needed to reproduce each of our experiments are documented in the Appendix in Table~\ref{tab:korg-inputs}.

\korg is limited by the level of detail in the model,
the definition of ``attack'' used by \korg~\cite{von2020automated},
and the \amodels and properties considered.
Thus
there could exist other attacks beyond those \korg synthesizes,
which violate other properties or work in other \amodels;
or attacks other than the type that \korg can find (e.g., statistical ones); 
or attacks that cannot be found without a more detailed model. 
These limitations are inherent to all attack synthesis techniques.

\subsection{\Amodels}
\label{subsec:amodels}
We use the term \emph{\amodel} to mean a formal description of the placement and capabilities of the attacker and protocol peer(s) on the network.
We create four \amodels: Off-Path, Evil-Server, Replay, and On-Path.
They are general-purpose and applicable to any transport protocol, and we contribute them to \korg.

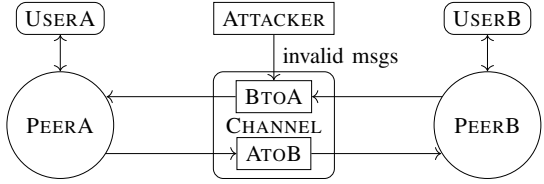
\begin{figure}
\begin{adjustbox}{width=0.4\textwidth,center}
\begin{tikzpicture}
\node[draw, rectangle, rounded corners, minimum width=1.7cm, minimum height=1.5cm, fill=white] (channel) at (3,0) 
    {\small \textsc{Channel}};
\node[draw,rectangle] (AtoB) at (3,-0.4) {\small \textsc{AtoB}};
\node[draw,rectangle] (BtoA) at (3,0.4) {\small \textsc{BtoA}};
\draw[->] (0,-0.4) to (AtoB);
\draw[->] (AtoB) to (5.35,-0.4);
\draw[->] (6,0.4) to (BtoA);
\draw[->] (BtoA) to (0.65,0.4);
\node[draw, circle, minimum height=1.5cm,fill=white] (sender) at (0,0) 
    {\small \textsc{PeerA}};
\node[draw, circle, minimum height=1.5cm,fill=white] (receiver) at (6,0) 
    {\small \textsc{PeerB}};
\node[draw, rectangle, fill=white] (attacker) at (3,1.5)
    {\small \textsc{Attacker}};
\draw[->] (attacker) to 
    node[right,text width=1.9cm] {\small invalid msgs}
(BtoA);
\node[draw,rectangle,rounded corners] (userA) at (0,1.5) {\small \textsc{UserA}};
\node[draw,rectangle,rounded corners] (userB) at (6,1.5) {\small \textsc{UserB}};
\draw[<->] (userA) to (sender);
\draw[<->] (userB) to (receiver);
\end{tikzpicture}
\end{adjustbox}
\caption{Off-Path \Amodel: $S =  \textsc{Attacker} \parallel \textsc{UserA} \parallel \textsc{PeerA} \parallel \textsc{Channel} \parallel \textsc{PeerB} \parallel \textsc{UserB}$.  The attacker can transmit messages into the BtoA buffer, but cannot receive messages, nor block messages in-transit.  The attacker can send only chunks having an invalid itag and vtag (as it is not privy to the association).}
\label{fig:offPathTM}
\end{figure}

\emph{Off-Path.}  In this model, an attacker who does not know either vtag communicates with one peer in order to disrupt the association formed by the two peers that want to communicate.
The vtag mechanism in SCTP was designed to defend against such an attacker.  
See Figure~\ref{fig:offPathTM}.

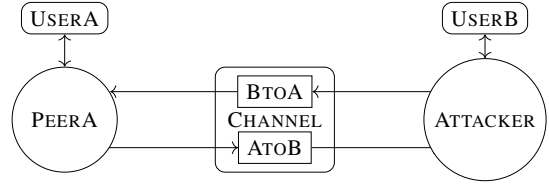
\begin{figure}
\begin{adjustbox}{width=0.4\textwidth,center}
\begin{tikzpicture}
\node[draw, rectangle, rounded corners, minimum width=1.7cm, minimum height=1.5cm, fill=white] (channel) at (3,0) 
    {\small \textsc{Channel}};
\node[draw,rectangle] (AtoB) at (3,-0.4) {\small \textsc{AtoB}};
\node[draw,rectangle] (BtoA) at (3,0.4) {\small \textsc{BtoA}};
\draw[->] (0,-0.4) to (AtoB);
\draw[->] (AtoB) to (5.35,-0.4);
\draw[->] (6,0.4) to (BtoA);
\draw[->] (BtoA) to (0.65,0.4);
\node[draw, circle, minimum height=1.5cm,fill=white] (sender) at (0,0) 
    {\small \textsc{PeerA}};
\node[draw, circle, minimum height=1.5cm,fill=white] (receiver) at (6,0) 
    {\small \textsc{Attacker}};
\node[draw,rectangle,rounded corners] (userA) at (0,1.45) {\small \textsc{UserA}};
\node[draw,rectangle,rounded corners] (userB) at (6,1.45) {\small \textsc{UserB}};
\draw[<->] (userA) to (sender);
\draw[<->] (userB) to (receiver);
\end{tikzpicture}
\end{adjustbox}
\caption{Evil-Server \Amodel: $S =  \textsc{UserA} \parallel \textsc{PeerA} \parallel \textsc{Channel} \parallel \textsc{Attacker} \parallel \textsc{UserB}$.  Peer B is prefixed with an attacker, whose code consists of a finite, terminating sequence of communication operations.}
\label{fig:evilServerTM}
\end{figure}

\emph{Evil-Server.}  In this model, one of the peers behaves maliciously. For example,  the attacker takes the form of a finite sequence of malicious instructions inserted before the code of Peer~B, after which B behaves like normal.  See Figure~\ref{fig:evilServerTM}.

\begin{figure}
\begin{adjustbox}{width=0.4\textwidth,center}
\begin{tikzpicture}
\node[draw, rectangle, rounded corners, minimum width=1.7cm, minimum height=1.5cm, fill=white] (channel) at (3,0) 
    {\small \textsc{Channel}};
\node[draw,rectangle] (AtoB) at (3,-0.4) {\small \textsc{AtoB}};
\node[draw,rectangle] (BtoA) at (3,0.4) {\small \textsc{BtoA}};
\draw[->] (0,-0.4) to (AtoB);
\draw[->] (AtoB) to (5.35,-0.4);
\draw[->] (6,0.4) to (BtoA);
\draw[->] (BtoA) to (0.65,0.4);
\node[draw, circle, minimum height=1.5cm,fill=white] (sender) at (0,0) 
    {\small \textsc{PeerA}};
\node[draw, circle, minimum height=1.5cm,fill=white] (receiver) at (6,0) 
    {\small \textsc{PeerB}};
\node[draw, rectangle, fill=white] (attacker) at (3,1.5)
    {\small \textsc{Attacker}};
\draw[<->] (attacker) to 
    node[right] {\small unmodified msgs}
(BtoA);
\node[draw,rectangle,rounded corners] (userA) at (0,1.5) {\small \textsc{UserA}};
\node[draw,rectangle,rounded corners] (userB) at (6,1.5) {\small \textsc{UserB}};
\draw[<->] (userA) to (sender);
\draw[<->] (userB) to (receiver);
\end{tikzpicture}
\end{adjustbox}
\caption{Replay \Amodel: $S =  \textsc{Attacker} \parallel \textsc{UserA} \parallel \textsc{PeerA} \parallel \textsc{Channel} \parallel \textsc{PeerB} \parallel \textsc{UserB}$. The attacker can capture and re-transmit messages in BtoA, but cannot edit captured messages, nor block those in transit.}
\label{fig:replayTM}
\end{figure}
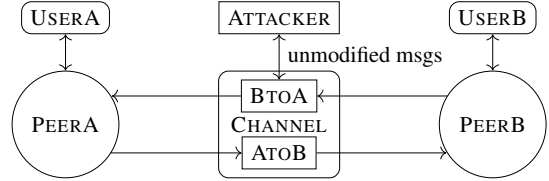

\emph{Replay.} In this model, the attacker can replay captured packets without modification. See Figure~\ref{fig:replayTM}.

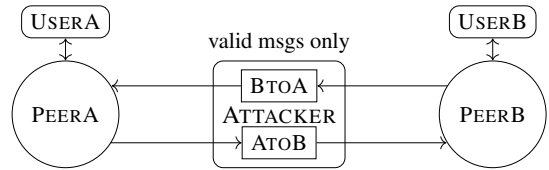
\begin{figure}
\begin{adjustbox}{width=0.4\textwidth,center}
\begin{tikzpicture}
\node[draw, rectangle, rounded corners, minimum width=1.7cm, minimum height=1.5cm, fill=white] (channel) at (3,0) 
    {\textsc{Attacker}};
\node[draw,rectangle] (AtoB) at (3,-0.4) {\small \textsc{AtoB}};
\node[draw,rectangle] (BtoA) at (3,0.4) {\small \textsc{BtoA}};
\draw[->] (0,-0.4) to (AtoB);
\draw[->] (AtoB) to (5.35,-0.4);
\draw[->] (6,0.4) to (BtoA);
\draw[->] (BtoA) to (0.65,0.4);
\node[draw, circle, minimum height=1.5cm,fill=white] (sender) at (0,0) 
    {\small \textsc{PeerA}};
\node[draw, circle, minimum height=1.5cm,fill=white] (receiver) at (6,0) 
    {\small \textsc{PeerB}};
\node[draw,rectangle,rounded corners] (userA) at (0,1.3) {\small \textsc{UserA}};
\node[draw,rectangle,rounded corners] (userB) at (6,1.3) {\small \textsc{UserB}};
\draw[<->] (userA) to (sender);
\draw[<->] (userB) to (receiver);
\node[] (valid) at (3,1) {\small valid msgs only};
\end{tikzpicture}
\end{adjustbox}
\caption{On-Path \Amodel: $S =  \textsc{UserA} \parallel \textsc{PeerA} \parallel \textsc{Attacker} \parallel \textsc{PeerB} \parallel \textsc{UserB}$.  The attacker is allowed to perform a finite sequence of send/receive actions, in which it only sends valid messages (but can receive anything).  Once this sequence terminates, it behaves like an honest channel.}
\label{fig:onPathTM}
\end{figure}

\emph{On-Path.} .  In this \amodel, the attacker controls the channel connecting the two peers, and can drop or insert valid messages at-will.  SCTP was not designed to provide security against such an attacker and we study this \amodel only to understand what the ``worst case scenario'' for SCTP looks like.  See Figure~\ref{fig:onPathTM}.

\subsection{Changes to \korg}

We improved \korg to support our SCTP analysis in four ways.
(1) Since \korg was originally hard-coded for enum-style packets, we extended \korg to support arbitrary finite packet types. 
This was needed to support our SCTP model (Figure~\ref{fig:bnf}).
(2) We modified \korg to report any attacks it finds even if it fails to exhaust the search-space.
Previously, it would report an error and discard any results if the space was not exhausted.
(3) To save time, 
we disabled the preliminary step where \korg verifies that the property holds in the absence of an attacker,
instead manually performing this step in \spin.
(4) We extended \korg to support support replay attackers.

A replay attacker is one capable of capturing and replaying messages. Although the replay attacker model reasons about packets received, the attacks this model produces form a subset of those produced by the \textit{On-Path} attacker; thus, soundness and completeness follow from the proofs in the \korg paper. Our replay attacker synthesis implementation supports packet capture and replay over the same or 
different channels. In the latter case, the attacker can capture a message from one channel and replay it into another.
It also supports packet storage in a memory buffer with configurable size, though,
the verification complexity increases exponentially with the memory bound.
Finally, to support the state change that happens when a new vtag is chosen,
we added a feature where a special message can be configured to flush the storage.

We also contribute our SCTP model and four \amodels in a format amenable to \korg.
We document the \amodels in Section~\ref{subsec:amodels}.
Excluding the models, 
our modifications required changing 80 lines of preexisting code and adding 213 lines of new functionality in \korg,
in addition to 102 lines of shell-script to automate our experiments.
All modifications are available with the paper artifacts.

\section{Experimental Results}\label{sec:results}

We next present our experimental results.
Our SCTP model 
satisfies all ten properties in the absence of an attacker.
To examine whether these properties still hold when an attacker is present,
we synthesize attacks using the three
\amodels. 
Then, we enable the CVE patch described in RFC 9260 and repeat our 
analysis, in order to check whether the patch resolves the vulnerability, 
and/or introduces any new attack vectors.
Finally, we show how a new attack is enabled if the ambiguous text identified in Section~\ref{subsec:ambiguity} is misinterpreted.
Analysis runtimes are related in Section~\ref{subsec:performance} in the appendix.

\subsection{Experimental Methodology}
Each time we run \korg, we ask it to synthesize $\leq 10$ attacks.  
In our experience, after the first ten, subsequent attacks tend to be repetitive, differing only by actions that do not impact the attack outcome.  
We configure \korg with a default search depth of 600,000, 
and a maximum depth of 2,400,000.
In our experience, these parameters balance fast-performance on smaller properties
with the ability to also attack more complex ones, without needing to run on a cluster.
We make certain assumptions or optimizations in the different \amodels.

\noindent~$\bullet$~\emph{Off-Path:} We assume the Off-Path attacker knows the port and IP of a peer, since otherwise, all its (spoofed) messages will be immediately discarded.\footnote{The ports and IP of a peer might not change between associations~\cite{stewart2008sctp}.} 
To reduce the search-space, we assume the attacker does not send \Data or \CookieError chunks,
which cannot change the receiving peer's state.
We further reduce the space by
first synthesizing attacks against the establishment routine,
where the attacker could only send messages that are used during establishment;
and then doing the same for teardown.
Our search is complete despite this split because the FSM is inherently Markovian and our properties do not look back more than one state in the past.
In our open-source artifacts, we provide code illustrating how this optimization can be repeated for any transport protocol.

\noindent~$\bullet$~\emph{Evil-Server:} We assume the Evil-Server attacker only sends valid messages, since it knows the current vtags.  To reduce the search-space, we assume it does not send \Data.

\noindent~$\bullet$~\emph{Replay:} We configure the replay attacker to have a memory size of two --- we more memory causes state-space explosion and makes exhaustive verification infeasible. We also configure it to discard all messages in memory upon receiving an \Init chunk, as this allows us to correctly model the vtag change between multiple connection and teardown cycles.

\noindent~$\bullet$~\emph{On-Path:} We perform the same optimizations as in the Off-Path \amodel.  And like in the Evil-Server \amodel, we assume the attacker only sends valid messages.

Our experiments are easily reproducible using the paper artifacts.
We summarize the inputs given to \korg for each experiment in Table~\ref{tab:korg-inputs} in the Appendix.

\subsection{Attacks}
We generate at least one attack in each \amodel,
all of which we summarize in Table~\ref{tab:synthesized-attacks}.
We discuss results for each \amodel in detail below.

\begin{table*}[]
\centering
\begin{tabular}{llp{13cm}}
Attacker Model               & Property & Synthesized Attacks                                                                                                                                                                                                                                              \\\hline
Off-Path                     & $\phi_9$ & A single variant of the CVE attack. \\\hline
\multirow{4}{*}{Evil-Server} & $\phi_1$ & One attack where the attacker guides A through passive establishment.  Then when A attempts active teardown, if its Shutdown Timer never fires, it deadlocks.                                                                                                     \\
                             & $\phi_6$ & One attack where the attacker guides A to \ShutdownReceived, then sends it an unexpected \CookieEcho, causing it to go back to \Established.                                                                         \\
                             & $\phi_8$ & One attack where the attacker guides A through most of active establishment before aborting the connection.  When the attack terminates, B receives the en-route \CookieEcho and completes passive establishment, creating a half-open connection. \\
                             & $\phi_9$ & One attack where the attacker guides A through passive establishment then terminates.  If B then attempts active establishment, the property fails, since the peers are de-synchronized. \\\hline 
Replay                       & $\phi_2$ & One attack where the attacker sends an \Abort before the peers establish a TSN for the association. \\\hline
\multirow{3}{*}{On-Path}     & $\phi_5$ & Four attacks where the attacker manipulates both peers into \ShutdownReceived.                                                                                                                                                                     \\
                             & $\phi_8$ & Two attacks where the attacker spoofs A to guide B through passive establishment.                                                                                                                                                                                 \\
                             & $\phi_9$ & Two attacks where the attacker spoofs B to guide A through passive establishment.                                                                       
\end{tabular}
\caption{Attacks found.}
\label{tab:synthesized-attacks}
\end{table*}

\textbf{Off-Path.}  \korg found a variant of the attack reported in CVE-2021-3772,
given in Figure~\ref{fig:cve-attack}. 
The variant differs only from the CVE in that it begins by transmitting some OOTB messages that are discarded and have no impact on the outcome.  It ends with the transmission of an \texttt{INIT} with an unexpected (zero) itag, which is the CVE attack. 

\begin{figure}
\[\begin{aligned}
   \textsc{BtoA} &\, ! \, \CookieAck,\texttt{U},\texttt{N}; & \text{(repeat twice more)} \\
   \textsc{BtoA} &\, ! \, \CookieEcho,\texttt{U},\texttt{N};\\
   \textsc{BtoA} &\, ! \, \CookieAck,\texttt{U},\texttt{N}; & \text{(repeat 6 more times)} \\
   \textsc{BtoA} &\, ! \, \CookieEcho,\texttt{U},\texttt{N};\\
   \textsc{BtoA} &\, ! \, \Init,\texttt{N},\texttt{U}; & \texttt{/* attack */}
\end{aligned}\]
\caption{Automatically synthesized CVE attack in the Off-Path \amodel.  \textsc{BtoA} is the channel from the attacker to the peer being attacked.  Only the final line matters.}
\label{fig:cve-attack}
\end{figure}

\textbf{Evil-Server.}  \korg synthesizes four attacks.  The first attack models a scenario in which the Shutdown Timer is configured to a very large value, and thus the attacker can cause a peer in active teardown to (essentially) deadlock by never responding to its \Shutdown message.  In the second attack, the attacker exploits the unexpected packet logic in $\mathsection$5.2.4 to guide a peer out of passive teardown and back into \Established.  The third and fourth attacks are similar 
and involve guiding one peer through establishment to de-synchronize it with the other, leading to a half-open connection.  

\textbf{Replay.} \korg synthesizes one attack where the attacker captures and replays an \Abort message sent by a peer before both peers establish a new TSN. The attacker can keep replaying the \Abort message indefinitely, preventing the peers from establishing a connection. No other attacks were found. This is expected, as the OOTB logic and TSNs should prevent a replay attacker from injecting old packets.  

\textbf{On-Path.}  \korg synthesizes four similar attacks where the attacker guides both peers into an association, and then spoofs each peer, sending a \Shutdown to the other.  In general, an On-Path attacker is so powerful that we expect it can manipulate either peer into any state it pleases, as it totally controls the network, so this is unsurprising.  \korg synthesizes two more attacks, one for each of the 
half-open properties, 
both similar to the last attack reported with the Evil-Server \amodel.

Note, the reason we do not rediscover the CVE attack in the Evil-Server or On-Path \amodel is that we restrict the attacker in both to only send valid messages, whereas the CVE attack requires an invalid \Init. We put this restriction in place in our model to avoid state-space explosion. In general, every attack that is possible in the Off-Path \amodel is also possible in the Evil-Server and On-Path ones.

\subsection{Patch Verification}

Next, we re-run our analysis with the CVE patch enabled.
In the Off-Path \amodel, \korg terminates without finding any attacks.
Since \korg found the vulnerability in the Off-Path \amodel when the patch was disabled,
and reports no attacks in that same \amodel when the patch is enabled,
and as \korg is complete,
this suffices to prove that the patch resolves the vulnerability.
In the other \amodels, we find the exact same attacks as those reported in Table~\ref{tab:synthesized-attacks}, and nothing more.
This proves the patch does not introduce any new attack vectors against our properties.

\subsection{Ambiguity Analysis}

\begin{figure}
\begin{adjustbox}{width=0.35\textwidth,center}
\begin{tikzpicture}[font=\sffamily,>=stealth',thick,
commentl/.style={text width=3cm, align=right},
commentr/.style={commentl, align=left},]
\node[] (init) {\small \textsc{Attacker}};
\node[right=1cm of init] (recv) {\small \textsc{PeerA}};
\node[right=1cm of recv] (third) {\small \textsc{PeerB}};
\node[below=0.1cm of recv] {\small \CookieEchoed};
\node[below=0.1cm of third] {\small \Closed};
\node[below=2.8cm of third] {\small \Closed};
\node[below=3cm of recv] {\small \Closed};
\draw[->] ([yshift=-0.8cm]init.south) 
            coordinate (fin1o) -- 
            ([yshift=-.2cm]fin1o-|recv) 
            coordinate (fin1e) 
            node[pos=.3, above, sloped] {\scriptsize \Init,itag=$i_3$};
\draw[->] ([yshift=-0.8cm]fin1o-|recv) 
            coordinate (fin1e) -- 
            ([yshift=-.2cm]fin1e-|third) 
            coordinate (fin2e) 
            node[pos=.3, above, sloped] {\scriptsize \InitAck,vtag=$i_1$,itag=$i_1$};
\draw[->] ([yshift=-0.8cm]fin2e-|third)
           coordinate (fin1w) --
           ([yshift=-.2cm]fin1w-|recv)
           coordinate (fin2w)
           node[pos=.3, above, sloped] {\scriptsize \Abort,vtag=$i_1$};
\draw[-] ([yshift=-0.8cm]recv.south)
          coordinate (hwtop) --
            ([yshift=-0.9cm,xshift=0.8cm]recv.south) --
            ([yshift=-1.1cm,xshift=0.8cm]recv.south);
\draw[-] ([yshift=-1.8cm,xshift=0.8cm]recv.south) --
         ([yshift=-2.2cm,xshift=0.8cm]recv.south);
\draw[->] ([yshift=-2.8cm,xshift=0.8cm]recv.south) --
          ([yshift=-3.5cm,xshift=0.8cm]recv.south) --
          ([yshift=-2cm]fin2e-|third)
          coordinate (final)
          node[pos=.9, above, sloped] {\scriptsize \CookieEcho,vtag=$i_2$};
\node[below=4cm of third] {\small \Closed};
\node[below=5cm of third] {\small \Established};
\node[below=5cm of recv] {\small \Closed};
\draw[->] ([yshift=-0.8cm]final-|third)
           coordinate (final2) --
           ([yshift=-.2cm]final2-|recv)
           coordinate (final3)
           node[pos=0.7, above, sloped] {\scriptsize \CookieAck,vtag=$i_1$};
\end{tikzpicture}
\end{adjustbox}
\caption{Message sequence chart showing the vtag-disclosure vulnerability enabled by misinterpretation of the ambiguous RFC text reported in Section~\ref{subsec:ambiguity}.  Note the strict timing requirements necessary for a successful attack.}
\label{fig:ambiguity-msc}
\end{figure}
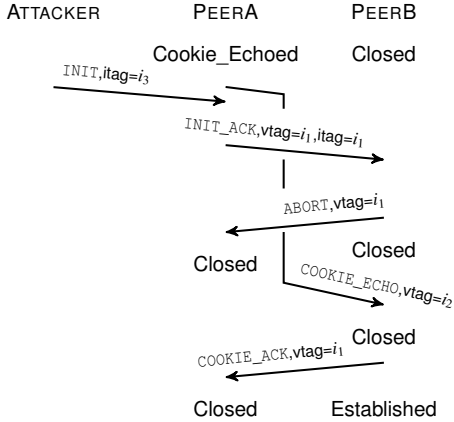

We next configure our model with the incorrect interpretation of the ambiguous text and run it interactively in \spin.
We find that the incorrect interpretation could enable a denial-of-service
in the form of a half-open connection, which we illustrate in Figure~\ref{fig:ambiguity-msc}.
We consulted with the lead SCTP RFC author who confirmed that the misinterpretation we describe could enable such an attack.  The attack is not possible if the text is interpreted correctly.
Out of concern that a real implementation might have misinterpreted the RFC document, we 
manually analyzed the source for both the Linux and FreeBSD implementations, 
and tested both implementations with \textsc{PacketDrill}, finding that neither made this mistake.

\section{Related Work}\label{sec:related}
There are many automated 
attack discovery tools,
each crafted to a particular variety of bug or mechanism of attack,
e.g., \textsc{SNAKE}~\cite{jero2019leveraging} 
(which fuzzes network protocols), 
\textsc{TCPwn}~\cite{jero2018automated} 
(which finds throughput attacks against TCP congestion control implementations),
\textsc{Tamarin}~\cite{meier2013tamarin} and \textsc{ProVerif}~\cite{blanchet2016modeling}
(which find attacks against secrecy in cryptographic protocols),
\korg~\cite{von2020automated} 
(which finds communication attacks against network protocols), and so on.
Some of these tools (e.g., \korg) 
are general purpose, designed to attack any correctness property,
while others (e.g., \textsc{Tamarin} or \textsc{ProVerif}) 
are designed to target specific types of properties such as secrecy and trace-equivalence.
One work, which studied TCP and ABP, suggested reactive controller synthesis (RCS) as an alternative to \korg's approach~\cite{matsui2022synthesis}.  \korg generates attacks that sometimes succeed, depending on choices made by the peers, whereas the RCS method only outputs attacks that always succeed; but such attacks do not always exist.
Another approach, which Fiterau-Brostean~et.~al.~\cite{fiterau2022automatabased}
successfully applied to various SSH~\cite{rfc4253} and DTLS~\cite{rfc_dtls} implementations,
describes incorrect behaviors using automata (rather than properties).
This specification style makes sense when generic bug patterns are known ahead of time.

Formal methods such as
theorem proving, model checking, property-based testing, and attack synthesis for protocols
have been applied to TLS~\cite{tls13_fm_ccs2017} and accountable proxying over it~\cite{bhargavan2018formal}, QUIC~\cite{quic_fm_sigcomm19}, Bluetooth~\cite{bluetooth_fm_sp2022}, 5G~\cite{5g_fm_ccs2018} and its key-establishment stack~\cite{miller20225g}, TCP congestion control~\cite{arun2021toward} and the combination of Karn's Algorithm and the RTO computation upon which it relies~\cite{von2023formal}, the TCP establishment routine~\cite{musuvathi2004model,von2020automated,pacheco2022automated}, and contactless EMV payments~\cite{basin2021emv,radu2022practical},
to name a few.
Compared to many of these systems,
such as TCP which has been studied for over 30 years~\cite{bellovin1989security,schuba1997analysis,harris1999tcp,jero2018automated,jero2019leveraging,von2020automated,pacheco2022automated,rfc_5927},
much less is known about the security of SCTP, particularly from an FM perspective.

Of the prior works that applied formal methods to the security of SCTP, 
only the Uppaal analysis by Saini and Fehnker~\cite{saini2017evaluating} used a technology 
(model-checking) that can verify arbitrary properties.
They reported two properties in their paper; the first is similar to our $\phi_2$.
The second says an adversary only capable of sending \Init packets cannot cause a victim peer to change state.  This property is trivial for us because we use an FSM model where the peer states are precisely the model states.
And in our model, the only transition out of \Closed that happens upon receiving an \Init is a self-loop that sends an \InitAck and returns to \Closed.  In contrast, in Saini and Fehnker's model the peer state is a variable in memory, while the model states are totally different (e.g., \textsf{LC1}, \textsf{LC2}).  Thus, the property merits verification in their model but not ours.
Saini and Fehnker's work is the only one we are aware of that 
studied SCTP in the context of an attacker using formal methods.
But their attacker was only capable of sending \Init messages,
in contrast to our \amodels which are much more sophisticated,
and their attacker could not spoof the port and IP of a peer.
Hence, they could not model (and so did not find) the CVE attack.

Another line of inquiry aims to model the performance of SCTP,
e.g., using numerical analyses and simulations~\cite{chukarin2006performance}.
For example, Fu and Atiquzzaman built an analytical model of SCTP congestion control,
including \emph{multihoming}, an SCTP feature not available in TCP.
They compared their model to simulations and found it to be accurate in estimating
steady-state throughput of multihomed paths~\cite{fu2005performance}.
Such models are also used to evaluate new features, e.g., as in~\cite{zou2006throughput}.

\section{Conclusion}\label{sec:conclusion}

In this work we formally modeled SCTP and specified ten novel LTL correctness properties 
based on a close reading of the RFCs.
We proved that in the absence of an attacker, 
    the protocol satisfies all ten properties.
We used \korg to synthesize attacks against our model
for four novel \amodels, 
    Off-Path, Evil-Server, Replay, and On-Path,
and for two configurations of the SCTP model -- one without the RFC 9260 patch and another with it.
This required improvements to \korg, which we open-sourced with the paper artifacts.
Without the patch, we found the CVE-2021-3772 attack in the Off-Path \amodel;
a variety of Evil-Server and On-Path attacks; and one Replay attack.
Then we repeated our analysis with the patch, and found that it eliminated the CVE vulnerability 
but did not eliminate any other attacks in other \amodels, nor introduce new vulnerabilities.

We also explored extending \korg to not just discover vulnerabilities, but synthesize patches too. We found the task infeasible as the search space for edits is enormous, and each edit requires re-verifying. And since \promela does composition over FIFO channels, reasoning about the composite Kripke Structure and tying it back to the \promela encoding proved very challenging.  Though we failed to synthesize patches in this work, we believe patch synthesis may be plausible in a more automata-theoretic context.

Our attacks highlight the need to explicitly handle unexpected but valid packets and set reasonable timer values.
We reported an ambiguity in RFC 9260, one interpretation of which could lead to a vulnerability.  We analyzed the Linux and FreeBSD SCTP implementations using \textsc{PacketDrill} and found both correctly interpreted the ambiguous text.  We concluded with a recommendation for how the text could be made unambiguous in an erratum or future RFC.

\bibliographystyle{acm}
\bibliography{main,rfcs}

\section{Appendix}\label{sec:appendix}
\subsection{History of CVE-2021-3772}
The vulnerability reported in CVE-2021-3772 arose from the following text in RFC 4960~\cite{rfc4960}:
\begin{quote}
If the value of the Initiate Tag in a received \Init chunk is found
to be 0, the receiver MUST treat it as an error and close the
association by transmitting an \Abort.
\end{quote}
As illustrated in Figure~\ref{fig:cve-msc}, if an implementation did not
check the validity of the \Init before transmitting an \Abort, then the RFC
allowed for a DoS attack where the attacker would transmit an invalid \Init
and thus trigger the victim to close an otherwise valid association.
This vulnerability was first reported in the SCTP mailing list~\cite{mailing}
and then reported in CVE-2021-3772~\cite{cve}.
The vulnerability was subsequently patched in the Linux implementation
by swapping the order of operations, to ensure the vtag is always checked before
the itag~\cite{linuxDiff}.
The RFC was updated in 9260~\cite{rfc9260} to say:
\begin{quote}
If the value of the Initiate Tag in a received \Init chunk is found to be 0, the receiver MUST silently discard the packet.
\end{quote} 
and FreeBSD~\cite{freebsd} implemented this patch when it was updated to reflect the new RFC.

\subsection{User Model}
We model not only each peer but also the user controlling each peer.  The user commands are \UserAssoc, \UserAbort, and \UserShutdown. A user and its peer are synchronously connected.  A subtlety of this composition is that the user is blocked from issuing unexpected commands, such as, a \UserAssoc when the peer is already in \Established.  When the user issues \UserAssoc or \UserShutdown the peer responds as shown in Figure~\ref{fig:fsm}.  When the user issues \UserAbort, the peer transmits an \Abort and transitions to \Closed.  The user is only allowed to issue a \UserAbort when the peer is in an active association.  The user herself is modeled nondeterministically with a single state, and a self-loop to and from that state to send each user command.

\subsection{Erratum to RFC 9260}
Beyond resolving the ambiguity described in Section~\ref{subsec:ambiguity}, we have several minor suggestions for places the SCTP RFC could be made more clear through an erratum.  First, we suggest incorporating the self-loop at \Closed that occurs upon receiving a \ShutdownComplete, into the State Association Diagram in Section 4, closing the active/active teardown routine.  Second, we suggest incorporating ``initialization collision'' (active/active establishment) into that same diagram, since it is a supported flow of the establishment routine and the default for WebRTC.  Third, we suggest expanding 5.2 to explain how to handle other unexpected chunks, e.g., \CookieError, \ShutdownComplete, etc., as such messages could be used by an Evil Server to deadlock a poorly implemented peer.

\subsection{Performance}\label{subsec:performance}
We time our experiments and patch verification tasks, and almost always, \korg terminates in seconds or minutes, with one interesting exception.
In the Off-Path experiments, \korg takes about two hours to confirm that no attacks exist against~$\phi_8$, and about an hour and a half to find the CVE attack against~$\phi_9$.  Recall that~$\phi_8$ and~$\phi_9$ are identical, except that the peer roles are reversed.  Further inspection reveals these two properties are the largest in our property set, and the Off-Path \amodel is the largest \amodel, as it includes four processes (two peers, an attacker, and a channel) whereas the others involve fewer.  The reason these two analyses take much longer than the others naturally follows, as \korg reduces to LTL model-checking, the runtime of which is polynomial in the size of the model and $\mathcal{O}(\log^2 |\phi|)$ in the size of~$\phi$~\cite{vardi1986automata}.
This is further highlighted by the fact that the patch verification tasks terminate in a few minutes -- and all the patch does is remove a transition from the code, reducing the model size.
We report all run-times in Table~\ref{tab:results-time}.

\begin{figure}
\begin{verbatim}
chan attacker_mem = [2] of { 
    mtype:msgs, 
    mtype:tag, 
    mtype:tag, 
    byte 
};
active proctype attacker_replay() {
  mtype:msgs b_0;
  mtype:tag b_1, b_2;
  byte b_3;
  do
  :: atomic { 
    AtoB ?? <b_0, b_1, b_2, b_3> 
    -> attacker_mem ! b_0, b_1, b_2, b_3; }
  :: atomic { 
    attacker_mem ?? b_0, b_1, b_2, b_3 
    -> AtoB ! b_0, b_1, b_2, b_3; }
  :: atomic { 
    attacker_mem ?? b_0, b_1, b_2, b_3; }
  :: break
  od
}
\end{verbatim}
\caption{The replay attacker gadget automatically synthesized by \korg.}
\end{figure}

\begin{table*}
\centering
\begin{tabular}{l|llllllll}
\multirow{2}{*}{} & \multicolumn{2}{c}{Off-Path} & 
                    \multicolumn{2}{c}{Evil-Server} &
                    \multicolumn{2}{c}{Replay} & 
                    \multicolumn{2}{c}{On-Path}\\
& E & P & E & P & E & P & E & P \\
                  \hline
$\phi_1$ & 2:20 & 
           2:13 & 
           0:23 & 
           0:23 &
           0:3 & 
           0:3 & 
           0:15 & 
           0:15 
           \\
$\phi_2$ & 8:43 & 
           11:14 & 
           0:21 & 
           0:21 &
           0:2 & 
           0:2 & 
           0:26 & 
           0:26 
           \\
$\phi_3$ & 3:20 & 
           12:53 & 
           0:20 & 
           0:20 &
           0:2 & 
           0:2 & 
           0:25 & 
           0:25 
           \\
$\phi_4$ & 1:45 & 
           1:26 &
           0:11 & 
           0:11 &
           0:2 & 
           0:2 & 
           0:14 & 
           0:14  
           \\
$\phi_5$ & 2:57 & 
           1:35 & 
           0:10 & 
           0:10 &
           0:2 & 
           0:2 & 
           0:12 & 
           0:12 
           \\
$\phi_6$ & 3:19 & 
           18:8 & 
           0:20 & 
           0:20 &
           0:2 & 
           0:2 & 
           0:25 & 
           0:25 
           \\
$\phi_7$ & 1:43 & 
           4:41 & 
           0:11 & 
           0:10 &
           0:2 & 
           0:2 & 
           0:13 & 
           0:14  
           \\
$\phi_8$ & 123:42 & 
           7:7 & 
           1:6 & 
           1:7 &
           0:2 & 
           0:2 & 
           1:34 & 
           1:34 
           \\
$\phi_9$ & 86:10 & 
           6:48 & 
           1:5 & 
           1:5 &
           0:2 & 
           0:2 & 
           0:11 & 
           0:11 
           \\
$\phi_{10}$ & 
0:4 & 
0:4 & 
0:3 & 
0:4 & 
0:2 & 
0:2 & 
0:4 & 
0:4 
\end{tabular}
\caption{Time taken (min:sec) to perform each (E) experiment and (P) patch verification on a 16GB M1 Macbook Air.}
\label{tab:results-time}
\end{table*}

\begin{figure*}
\begin{verbatim}
// Property 1
G((st[0] == Closed) -> (X(F(st[0] == Closed || st[0] == Established || st[0] == CookieWait))))
// Property 2
G(F(st[0] != ost[0] || st[1] != ost[1] || (st[0] == Closed && st[1] == Closed) || 
    (st[0] == Established && st[1] == Established)))
// Property 3
G((st[0] != ost[0] && ost[0] == ShutdownAckSent) -> (st[0] == Closed))
// Property 4
G(F(st[0] != CookieEchoed || timers[0] == T1_COOKIE))
// Property 5
G(st[0] != ShutdownReceived || st[1] != ShutdownReceived)
// Property 6
G((st[0] != ost[0] && ost[0] == ShutdownReceived) -> (st[0] == ShutdownAckSent || st[0] == Closed))
// Property 7
G(st[0] != CookieEchoed || st[1] != ShutdownReceived)
// Property 8
G((ost[1] == Established && ost[0] == Closed && everAborted == false && everTimedOut == false && 
    ost[0] != st[0]) -> (st[0] == Established || st[0] == IntermediaryCookieWait))
// Property 9
G((ost[0] == Established && ost[1] == Closed && everAborted == false && everTimedOut == false && 
    ost[1] != st[1]) -> (st[1] == Established || st[1] == IntermediaryCookieWait))
// Property 10
G((ost[0] == Established && (st[0] == ShutdownSent || st[0] == ShutdownReceived)) -> F(st[0] == Closed))
\end{verbatim}
\caption{Our ten LTL properties are formulated in \textsc{Promela} as follows.  We define our atomic propositions as follows in \promela, where \texttt{st} holds the state of each peer, \texttt{ost} holds the prior one, and \texttt{timers} holds the peers' timers.}
\end{figure*}

\begin{table*}[]
\centering
\begin{tabular}{llll}
Threat Model & $P$                                                                                                                    & $Q$                & I/O                               \\\hline
Off-Path     & $\textsc{UserA} \parallel \textsc{PeerA} \parallel \textsc{Channel} \parallel \textsc{PeerB} \parallel \textsc{UserB}$ & Empty Program    & Sends invalid messages            \\
Evil-Server  & $\textsc{UserA} \parallel \textsc{PeerA} \parallel \textsc{Channel} \parallel \textsc{UserB}$                          & $\textsc{PeerB}$   & Sends and receives valid messages \\
Replay  & $\textsc{UserA} \parallel \textsc{PeerA} \parallel \textsc{Channel} \parallel \textsc{PeerB} \parallel \textsc{UserB}$  & $\textsc{Channel}$   & Sends messages previously received \\
On-Path      & $\textsc{UserA} \parallel \textsc{PeerA} \parallel \textsc{PeerB} \parallel \textsc{UserB}$                            & $\textsc{Channel}$ & Sends and receives valid messages
\end{tabular}
\caption{\korg inputs for each \amodel.  $P$ is the invariant component, which the attacker cannot change.  $Q$ is the variant component, which the attacker can prefix with a finite sequence of communication events.  In the Off-Path \amodel, $Q$ is the empty program, because the Off-Path attacker does not attach itself to any pre-existing system component.  I/O specifies what the attacker is allowed to receive or send.  $\phi$ will be one of the properties in $\mathsection$\ref{subsec:properties}, and the peer model (run by \textsc{PeerA} and \textsc{PeerB}) is described in Section~\ref{sub:Model Structure}.}
\label{tab:korg-inputs}
\end{table*}

\end{document}